\input harvmac.tex

\input epsf.tex


\def\figin{\epsfcheck\figin}\def\figins{\epsfcheck\figins}
\def\epsfcheck{\ifx\epsfbox\UnDeFiNeD
\message{(NO epsf.tex, FIGURES WILL BE IGNORED)}
\gdef\figin##1{\vskip2in}\gdef\figins##1{\hskip.5in}
\else\message{(FIGURES WILL BE INCLUDED)}%
\gdef\figin##1{##1}\gdef\figins##1{##1}\fi}
\def\DefWarn#1{}
\def\figinsert{\goodbreak\midinsert}
\def\ifig#1#2#3{\DefWarn#1\xdef#1{fig.~\the\figno}
\writedef{#1\leftbracket fig.\noexpand~\the\figno}%
\figinsert\figin{\centerline{#3}}\medskip\centerline{\vbox{\baselineskip12pt
\advance\hsize by -1truein\noindent\footnotefont{\bf
Fig.~\the\figno:} #2}}
\bigskip\endinsert\global\advance\figno by1}



\lref\jmcn{
  J.~M.~Maldacena and C.~Nunez,
  Int.\ J.\ Mod.\ Phys.\  A {\bf 16}, 822 (2001)
  [arXiv:hep-th/0007018].
}

\lref\LLM{
  H.~Lin, O.~Lunin and J.~M.~Maldacena,
  JHEP {\bf 0410}, 025 (2004)
  [arXiv:hep-th/0409174].
}
\lref\hljm{
  H.~Lin and J.~M.~Maldacena,
  Phys.\ Rev.\  D {\bf 74}, 084014 (2006)
  [arXiv:hep-th/0509235].
}

\lref\FStwo{
  A.~Fayyazuddin and D.~J.~Smith,
  JHEP {\bf 0010}, 023 (2000)
  [arXiv:hep-th/0006060].
}

\lref\ward{
R.~S.~Ward,
Class.\ Quant.\ Grav.\  {\bf 7}, L95 (1990).
}

\lref\RS{   L.~Randall and R.~Sundrum,
  Phys.\ Rev.\ Lett.\  {\bf 83}, 3370 (1999)
  [arXiv:hep-ph/9905221].
}

\lref\wittenM{ E.~Witten,
  Nucl.\ Phys.\  B {\bf 500}, 3 (1997)
  [arXiv:hep-th/9703166].
}

\lref\pans{  P.~C.~Argyres and N.~Seiberg,
  JHEP {\bf 0712}, 088 (2007)
  [arXiv:0711.0054 [hep-th]].
}

\lref\GreenDI{
  M.~B.~Green and P.~Vanhove,
  Phys.\ Lett.\  B {\bf 408}, 122 (1997)
  [arXiv:hep-th/9704145].
}

\lref\HarveyBX{
  J.~A.~Harvey, R.~Minasian and G.~W.~Moore,
  JHEP {\bf 9809}, 004 (1998)
  [arXiv:hep-th/9808060].
}

\lref\BlauVZ{
  M.~Blau, K.~S.~Narain and E.~Gava,
  JHEP {\bf 9909}, 018 (1999)
  [arXiv:hep-th/9904179].
}

\lref\AharonyRZ{
  O.~Aharony, J.~Pawelczyk, S.~Theisen and S.~Yankielowicz,
  Phys.\ Rev.\  D {\bf 60}, 066001 (1999)
  [arXiv:hep-th/9901134].
}

      \lref\BilalPH{
        A.~Bilal and C.~S.~Chu,
        Nucl.\ Phys.\  B {\bf 562}, 181 (1999)
        [arXiv:hep-th/9907106].
      }

\lref\AharonyDJ{
  O.~Aharony and Y.~Tachikawa,
  JHEP {\bf 0801}, 037 (2008)
  [arXiv:0711.4532 [hep-th]].
}

\lref\NojiriMH{
  S.~Nojiri and S.~D.~Odintsov,
  Int.\ J.\ Mod.\ Phys.\  A {\bf 15}, 413 (2000)
  [arXiv:hep-th/9903033].
}
\lref\Kats{
  Y.~Kats and P.~Petrov,
  JHEP {\bf 0901}, 044 (2009)
  [arXiv:0712.0743 [hep-th]].
}

\lref\BachasUM{
  C.~P.~Bachas, P.~Bain and M.~B.~Green,
  JHEP {\bf 9905}, 011 (1999)
  [arXiv:hep-th/9903210].
}

\lref\wittenquantization{
  E.~Witten,
  J.\ Geom.\ Phys.\  {\bf 22}, 1 (1997)
  [arXiv:hep-th/9609122].
}

\lref\myers{
  R.~C.~Myers,
  JHEP {\bf 9912}, 022 (1999)
  [arXiv:hep-th/9910053].
}
\lref\KS{
  I.~R.~Klebanov and M.~J.~Strassler,
  JHEP {\bf 0008}, 052 (2000)
  [arXiv:hep-th/0007191].
}

\lref\Gaiotto{
  D.~Gaiotto,
  arXiv:0904.2715 [hep-th].
}

\lref\GMNtwo{ D. Gaiotto, G. Moore and A. Neitzke, to appear. }

\lref\AcharyaMU{
  B.~S.~Acharya, J.~P.~Gauntlett and N.~Kim,
  Phys.\ Rev.\  D {\bf 63}, 106003 (2001)
  [arXiv:hep-th/0011190].
}

\lref\HenningsonGX{
  M.~Henningson and K.~Skenderis,
  JHEP {\bf 9807}, 023 (1998)
  [arXiv:hep-th/9806087].
}

\lref\Pestun{
  V.~Pestun,
  arXiv:0712.2824 [hep-th].
}

\lref\GV{
  R.~Gopakumar and C.~Vafa,
  arXiv:hep-th/9812127.
}

\lref\deconstruction{
  N.~Arkani-Hamed, A.~G.~Cohen, D.~B.~Kaplan, A.~Karch and L.~Motl,
  JHEP {\bf 0301}, 083 (2003)
  [arXiv:hep-th/0110146].
}

\lref\skes{
  S.~Kachru and E.~Silverstein,
  Phys.\ Rev.\ Lett.\  {\bf 80}, 4855 (1998)
  [arXiv:hep-th/9802183].
}
\lref\mdgm{
  M.~R.~Douglas and G.~W.~Moore,
  arXiv:hep-th/9603167.
}

\lref\GauntlettNG{
  J.~P.~Gauntlett, N.~Kim and D.~Waldram,
  Phys.\ Rev.\  D {\bf 63}, 126001 (2001)
  [arXiv:hep-th/0012195].
}

\lref\PerniciNW{
  M.~Pernici and E.~Sezgin,
  Class.\ Quant.\ Grav.\  {\bf 2}, 673 (1985).
}

\Title{\vbox{\baselineskip12pt \hbox{} \hbox{
} }} {\vbox{\centerline{ The gravity duals of}
\vskip .7cm
\centerline{ ${\cal N}=2$ superconformal field theories
  }
\centerline{
  }
}}
\bigskip
\centerline{Davide Gaiotto and Juan Maldacena }
\bigskip
\centerline{ \it  School of Natural Sciences, Institute for
Advanced Study} \centerline{\it Princeton, NJ 08540, USA}

\vskip .3in \noindent
  We study the gauge/gravity duality for theories with four dimensional ${\cal N}=2$
  supersymmetries. We consider the large class of generalized quiver field
  theories constructed recently by one of us (D.G.).
   These field theories can also be viewed as the IR limit of M5 branes
   wrapping a Riemann surface with punctures.  We give a prescription for
   constructing the corresponding geometries and we discuss a few special cases in
   detail.
  There is a precise match for various quantities between the field theory
  and the M-theory description.


 \Date{ }


\newsec{ Introduction}

A large and interesting class of four dimensional
${\cal N}=2$
superconformal
field theories was recently constructed in \Gaiotto . These are constructed
as generalized quivers involving elementary fields as well as strongly coupled
field theories as building blocks.   These field theories
can also be viewed as coming from $N$ M5 branes that are wrapping a
Riemann surface, which could be a sphere, a torus, or a higher genus
surface. In addition we can have extra non-compact branes that
intersect this surface at points. These can be viewed as
``punctures'' on the Riemann surface. In this paper we examine these
constructions in the large $N$ limit where we can study the system
via its gravity dual.  This gives a large class of $AdS_5$ compactifications
of M-theory with four dimensional ${\cal N}=2$ supersymmetry which fall into
the ansatz proposed in \LLM . We make a precise correspondence between the
different field theories and the different gravity solutions.

A particular theory constructed in \Gaiotto\ is a strongly coupled
superconformal theory $T_N$ with three $SU(N)$ global symmetries.
The theory has no coupling constant, and comes from $N$ M5 branes
wrapping a three-punctured sphere, or  a ``pair of pants'' surface.
Using this theory as an a kind of three vertex, one can construct
quivers which have the topology of higher genus Riemann surfaces.
The corresponding gravity solutions are given by certain $AdS_5$
compactifications of M-theory whose internal manifold involves the
same Riemann surface with the same topology \jmcn .

The construction in \Gaiotto\ relies on new phenomena that happen
in quiver gauge theories when the original gauge couplings become
strong. The complex structure moduli of the Riemann surface,
including the position of the punctures, encode the gauge couplings
of the theory. Ordinary quiver gauge theories correspond to the simplest
  choice of punctures on a sphere. The strong coupling limit
brings several simple punctures together to produce more complicated
punctures.

By starting with these four dimensional quiver theories it is also possible
 to recover the six dimensional field theory by taking a limit where the number
 of nodes in the quiver goes to infinity. This is in the spirit of \deconstruction , and
 it gives some insight into the origin of the $N^3$ degrees of freedom for a fivebrane.

In \Gaiotto\ it was found that the possible punctures leading to
four dimensional superconformal field theories have a classification
in terms of Young diagrams of $SU(N)$. We discuss the gravity
solution that corresponds to each of these punctures. Some types of
punctures lead to non-abelian global symmetries. In the
bulk, these arise from $A_{k-1}$ singularities. It is interesting that the five
dimensional gauge coupling of these gauge fields can be fairly
strong, which leads to global symmetries with two point functions
for the current of order one. Theories with such global symmetries
might be useful  for model building.

Finally, we will give a general prescription for constructing the
gravity duals for general field theories in \Gaiotto\ by using the
general ansatz for gravity duals of theories with ${\cal N}=2 $ SUSY
written in \LLM . In fact, it is interesting to study and classify various
compactifications of M theory. Here we are studying compactifications of M-theory
to $AdS_5$ which preserve sixteen supercharges. Since we are in $AdS_5$ the superalgebra
is  the ${\cal N}=2$ superconformal algebra in four dimensions. We will find that
we can think of a large class of such compactifications in terms of very simple elementary
building blocks. We start with a Riemann surface and we put a variety of punctures on the
Riemann surface. This Riemann surface and the punctures give some boundary conditions for
a certain equation, whose solution  determines
the geometry.

  This paper is organized as follows. In section two we describe the theories
  that arise from M5 branes wrapping a compact Riemann surface of genus $g>1$.
  We start by reviewing a purely four dimensional construction of the corresponding
  field theory. In section three we discuss the possible punctures. Starting from the
  simplest ones and ending with the most general punctures. In section four we discuss
  some aspects of punctures which give rise to non-abelian global symmetries.
  In section five we conclude and discuss some open problems.

\newsec{ Field theories with no flavor symmetries and their gravity solutions }

In this section we construct the field theories for the geometries
discussed in \jmcn . Those geometries were obtained as the IR
limit of M5 branes wrapping a two dimensional Riemann surface with
a twisting that preserves four dimensional  ${\cal N}=2$
supersymmetry.
First we pause to review some aspects of the field theories discussed in \Gaiotto .

\subsec{Review of the construction of $T_N$ and the generalized quiver  field theories}

 \ifig\dfournsfive{ (a) A brane construction that leads to the quiver shown in (b).
 The circles are gauge groups, the squares are global symmetries and the links are bifundamentals. We have
 $N$ NS 5 branes in (a) and $N-1$ gauge groups in (b).
 The links that join a square to a circle are fundamentals of the gauge group and of the global symmetry.
 (c) The space of couplings of the theory is parametrized by the position of punctures on a sphere, up to $SL(2,C)$
 transformations.  There are two special punctures associated
 to the two $SU(N)$ global symmetries. We also have elementary  punctures associated to the NS fivebranes in (a).
   In (d) we schematically show the region in parameter space which
 leads to weak  coupling in the quiver in (b).
 } {\epsfxsize4in\epsfbox{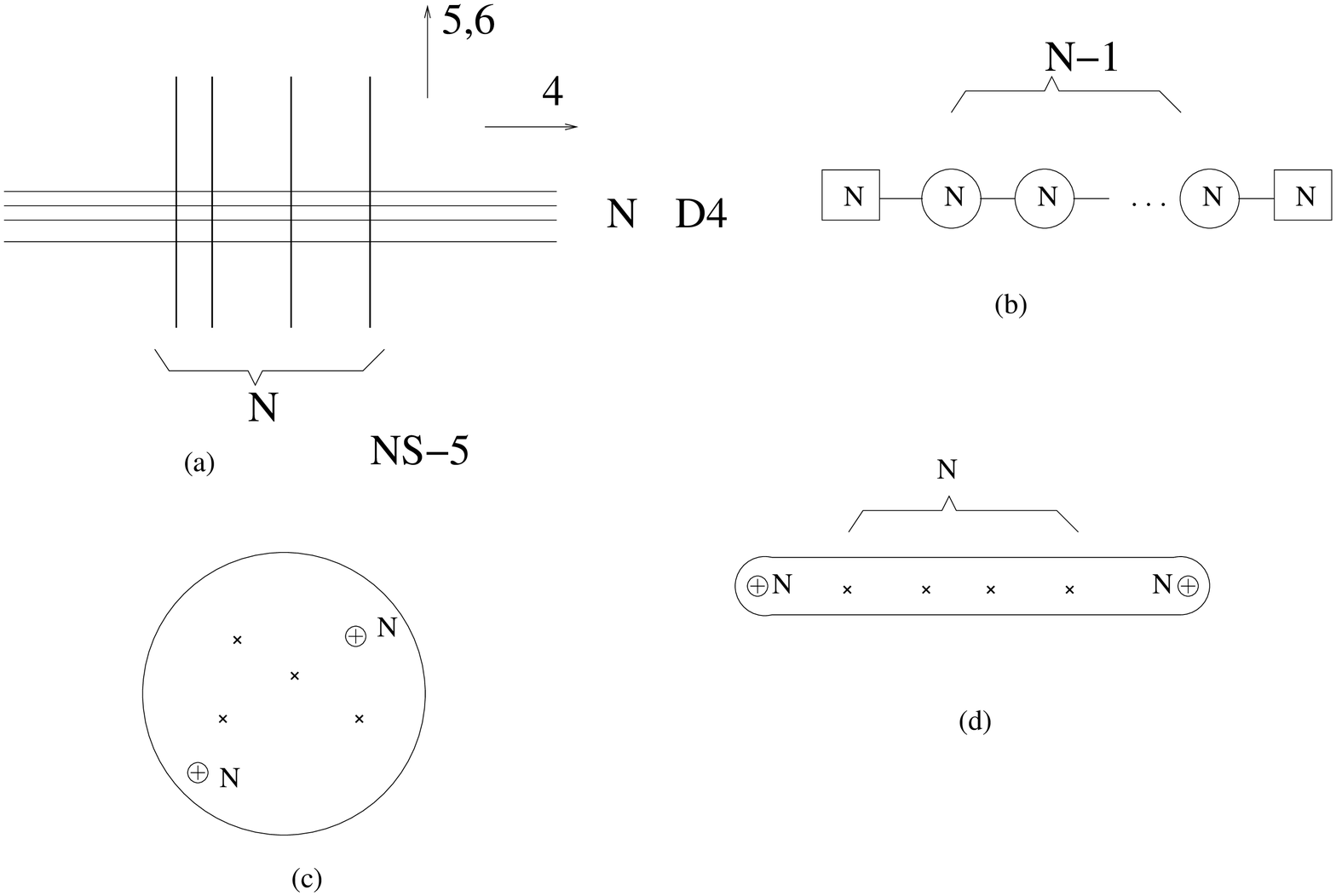} }

We   start by reviewing the construction of a particular ${\cal
N}=2$ superconformal field theory $T_N$ given in \Gaiotto. The
starting point is a quiver gauge theory that can be obtained by
considering $N$ D4 branes extended along the directions 01234 and
$N$ NS-5 branes along the directions 0123~56, see \dfournsfive (a).
This is a conformal quiver gauge theory with $N-1$ SU(N) gauge
groups. It has $N$ fundamentals at each of the two ends and
bifundamentals between consecutive nodes, see \dfournsfive (b). This
theory has $N-1$ coupling constants. At weak coupling we can take
them to be the $\tau_i$ parameters of each gauge group. Witten has
considered the M theory lift of this brane configuration \wittenM .
He argued that the couplings for this theory can be viewed as points
on a sphere, see \dfournsfive (c) \foot{One should not  confuse the Riemann surface with punctures
appearing in \dfournsfive (c) with the
Riemann surface that describes the Seiberg Witten curve. The latter is roughly an $N^{\rm th}$ cover of the first. The
latter encodes the couplings for the low energy theory while the first (the one in \dfournsfive (c)) describes
the couplings of the full UV theory.}.
 There are $N$ points associated to the $N$ NS-5branes
in the above picture and two special points associated to the
semi-infinite D4 branes. Thus we have $N+2$ punctures on a sphere.
The $N-1$ gauge couplings of the quiver are related to the $N-1$
cross ratios that we can make from these $N+2$ points.
 \ifig\quiver{ (a) The initial quiver associated to the brane construction in   \dfournsfive . We
 have quark fields $q$ at the ends  and bifundamental fields  $A$ between the gauge groups.
  There are $N-1$ gauge groups and $N-2$
 bifundamentals.
 In (b) we see a dual description of the same theory.
  The theory consists of the theory $T_N$ coupled to a quiver with decreasing ranks.
 In some region in parameter space the theory becomes
 weakly coupled.
 (c) If we decouple the $SU(N)$ factor in (b) we are left with the nontrivial theory $T_N$.
 The boxes denote global symmetries. The circles denote gauge groups. In addition, we have a global
 $U(1)$ symmetry for every link in the diagram.
  } {\epsfxsize2.3in\epsfbox{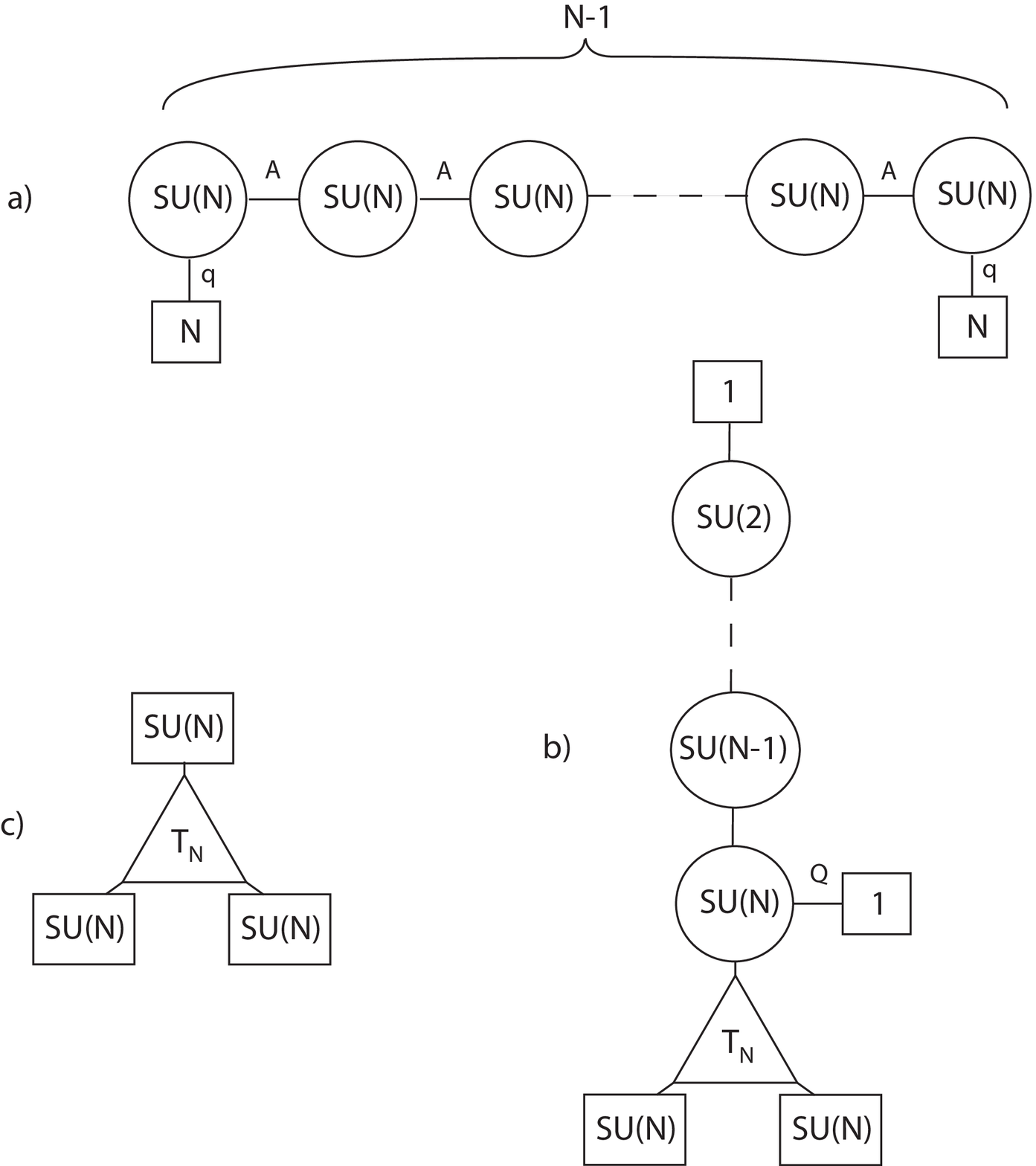} }

The quiver gauge theory has $N$ $U(1)$ global symmetry groups, and
two $SU(N)$ global symmetry groups. In a precise sense, the two
$SU(N)$ flavor groups are associated to the two special punctures,
and the $N$ $U(1)$ flavor groups to the  other $N$ punctures. The
weak coupling limit of any node of the original quiver corresponds to
a situation where   some of the $N$ punctures are brought towards
one special puncture, the others towards the other special puncture.
It was argued in \Gaiotto\ that other ways to bring punctures
together in a hierarchical fashion, which correspond to strong
coupling limits of the original quiver, produce alternative weakly
coupled descriptions of the theory via a generalization of the
Argyres-Seiberg duality \pans. Namely, the theory has hidden dual
gauge groups which are becoming weakly coupled, but a piece of the
theory remains strongly coupled. In the simple situation where the
two special punctures are brought together, a new dual $SU(N)$ gauge
group emerges. The theory is now described by a generalized quiver
as in \quiver (b). This new quiver looks like an ordinary quiver
except for the $T_N$ factor. The $T_N$ theory is an ${\cal N}=2$
superconformal field theory with three $SU(N)$ flavor symmetries and
no coupling constant. In this case, one of these three $SU(N)$
factors is gauged, the other two are the $SU(N)$ flavor symmetries
associated to the original two special punctures. The contribution
to the $SU(N)$ beta function from the $T_N$ theory is the same as
that of $N$ fundamental hypers. Thus, we can see that all the nodes
in the quiver in  \quiver\   are conformal.

In the limit that we take the $SU(N)$ gauge coupling to zero we
decouple the $T_N$ theory. Thus, we can consider this theory on its
own. In the decoupling limit the $N+2$ punctured sphere simply
degenerates to a three punctured sphere, representing $T_N$, and a
sphere with $N+1$ punctures, accounting for the remaining ``tail''
quiver. The two new punctures produced by the degeneration limit are
identical to the two original special punctures. Viceversa, later we
will introduce new $SU(N)$ gauge groups to glue together several
three-punctured spheres into any desired Riemann surface, replacing
pairs of special punctures with handles.

Let us discuss now some of the properties of this theory. First,
it is a theory with no coupling constant. We can compute its
conformal anomaly $a$ and $c$ coefficients as follows. We first
start from the usual formulas for $a$ and $c$ for weakly coupled
${\cal N}=2$ conformal theories \eqn\ntwo{ \eqalign{
c = &  { 2 n_v + n_h \over
12 } ~,~~~~~a = { 5 n_v + n_h \over 24} ~,~~~~  24 (a-c) =     n_v - n_h
\cr
Tr[U(1)_R] = Tr[U(1)_R^3] =&  2(n_v-n_h) ~,~~~~~~~~Tr[ U(1)_R J^3_R J^3_R] = { 1 \over 2 } n_v
}
} where $J^3_R$ is the $J^3$ generator in $SU(2)_R$\foot{ $U(1)_R$ is the ${\cal N}=2$ $U(1)_R$ charge
normalized so that the supercharge has charge one. $J^3_R$ has eigenvalues $\pm 1/2$ on the fundamental
of $SU(2)_R$.}. We interpret \ntwo\ in two ways, first we can view it as the computations of $a,c$ in
a free theory. We can also view them as the definition of effective numbers $n_v$ and $n_h$ for an arbitrary
theory. In fact, instead of computing $a$ and $c$ we will compute the effective numbers $n_v$ and $n_h$ defined
through \ntwo . Finally the second line of \ntwo\ is simply stating which anomalies can be used to
compute these two quantities.
We start with the values of $a$ and $c$ for the
original quiver and then we subtract the contributions from the
weakly coupled chain in \quiver (b). We find \eqn\orig{ n_v^{\rm orginal } =
(N^2-1) (N-1) = N^3 -N^2 -N +1 ~,~~~~~~~~~~~~n^{\rm original }_h =
N^3 } \eqn\chain{\eqalign{ n_v^{\rm chain} &= \sum_{i=2}^{N} (i^2
-1) = { N^3 \over 3} + {N^2 \over 2 } - { 5 N \over 6 } \cr
n_h^{\rm chain} & = 2 + N +  \sum_{i=2}^{N-1} i(i+1 ) = {N^3 \over
3 } + { 2 N \over 3 } }}

Thus we conclude that \eqn\tnval{ \eqalign{ n_v^{T_N} &=  { 2 N^3
\over 3 } - { 3 N^2 \over 2 } - { N \over 6 } + 1 \cr n_h^{T_N} & =
{ 2 N^3 \over 3 } - { 2 N  \over 3} }} These are not the actual
numbers of fields. It is simply a way to parametrize $a$ and $c$,
and the anomaly coefficients \ntwo .

This theory contains an interesting BPS operator which is constructed as
follows. In the original quiver theory we take
 a quark, a sequence of bifundamentals and an
antiquark in such a way that we go from one end of the quiver to the
other, see \quiver a. This operator has the schematic form $ H_{ij}
= q_i A_{1} A_2 \cdots A_{N-2} \tilde  q_j $. This operator is
protected, it has $SU(2)_R$ charge $j = N/2$ and zero $U(1)_R$
charge. Thus it has conformal dimension $\Delta = N$. It transforms
in the fundamental representation of both $SU(N)$ flavor groups, and
is also charged under all the $N$ $U(1)$ flavor symmetry groups. In
the dual quiver representation, the $N$ $U(1)$s are remixed into the
$U(1)$ flavor symmetry of the new bifundamentals, of the $SU(2)$
fundamental field and of the fundamental $Q^k$ that is attached
to the $SU(N)$ gauge group in \quiver b.

We conjecture a decomposition of the form $ H_{ij} =  O_{ijk} Q^k$
where $Q^k$ is the fundamental that is attached to the $SU(N)$ gauge
group in \quiver (b) and $O_{ijk}$ is an operator in the theory $T_N$.
 In the limit that we decouple the $SU(N)$ factor
we see that we get a new gauge invariant BPS operator given by
$O_{ijk}$. This operator transforms under the three $SU(N)$ groups
of the theory $T_N$ and it has dimension $\Delta = N-1$. We see that
something special happens for $N=2$. In this case $Q_{ijk}$ is just
a free field, leading to four free hypermultiplets. This is all the
theory $T_2$ contains. \quiver a and \quiver b both coincide with
the $N_f=4$ $SU(2)$ gauge theory, in two S-dual frames \Gaiotto . In the case
$N=3$ $O_{ijk }$ has dimension two and it contains a conserved
current in its multiplet. This  enhances the global symmetry from
$SU(3)^3$ to $E_6$. Indeed $T_3$ coincides with the same strongly
interacting SCFT with $E_6$ flavor symmetry which plays a crucial
role in \pans.

\ifig\quiverother{ An alternative way of producing the theory $T_N$. (a) The initial quiver
has $N-2$ gauge groups. (b) Dual description of the same theory. In this case we only gauge an $SU(N-1)$
subgroup of one of the $SU(N)$ symmetries of $T_N$.
  } {\epsfxsize3.0in\epsfbox{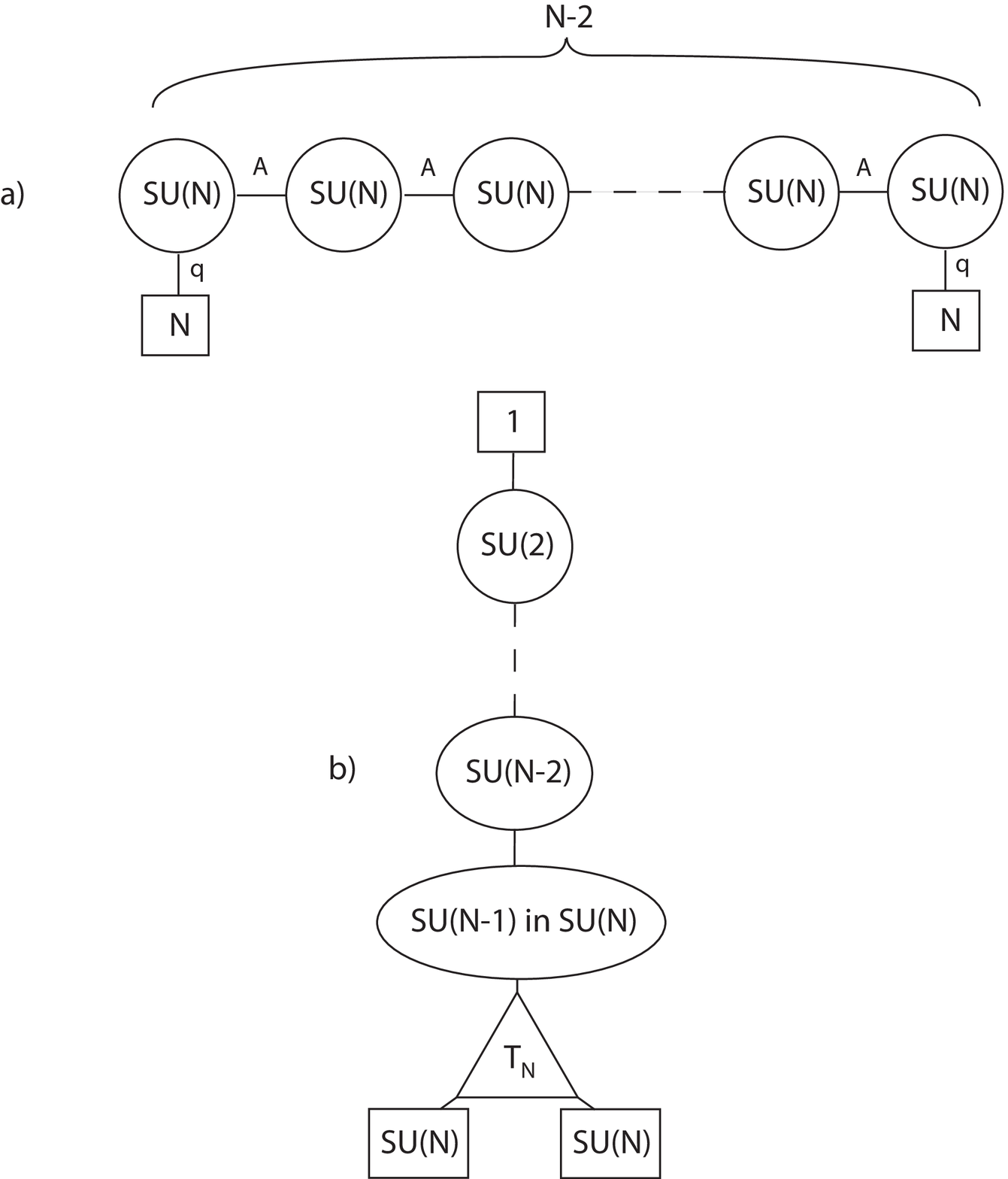} }

For higher values of $N$, $N>3$, we expect simply the $SU(N)^3$
symmetry, and no further global symmetries. The conjectured
existence of the operator $O_{ijk}$ can be confirmed by looking at a slightly
different setup: a quiver of the form of \quiver a, but one less
$SU(N)$ gauge node, see \quiverother (a). In the appropriate strong coupling this quiver
has a dual quiver description involving again $T_N$. The dual quiver
description resembles \quiver b, but the $SU(N)$ gauge group is
missing, and the $SU(N-1)$ gauge group directly gauges the simple
$SU(N-1)$ subgroup of a $SU(N)$ global symmetry group of $T_N$, see \quiverother (b). The
current anomalies again allow this gauge group to be conformal. Now
$H_{ij}$ has dimension $N-1$ and simply coincides with $O_{ijN}$,
the piece of $O_{ijk}$ which is a singlet of the $SU(N-1)$ acting on
$k=1 \cdots N-1$. It is easy to show that any alternative hypothesis
involving the remaining bifundamentals and quarks of the \quiver b
in the decomposition of $H_{ij}$ would produce several distinct
operators of the form $H_{ij}$, but only one is present in the
original quiver.
\ifig\symmetric{ (a) Starting initial quiver where we attach chains to the ends of the quiver in \quiverother .
(b) This theory has this alternative description, using the results in \quiverother . In this case the three
vertices of the $T_N$ theory are treated in the same way. Thus the configuration is invariant under permutations
of the three chains.
  } {\epsfxsize4.0in\epsfbox{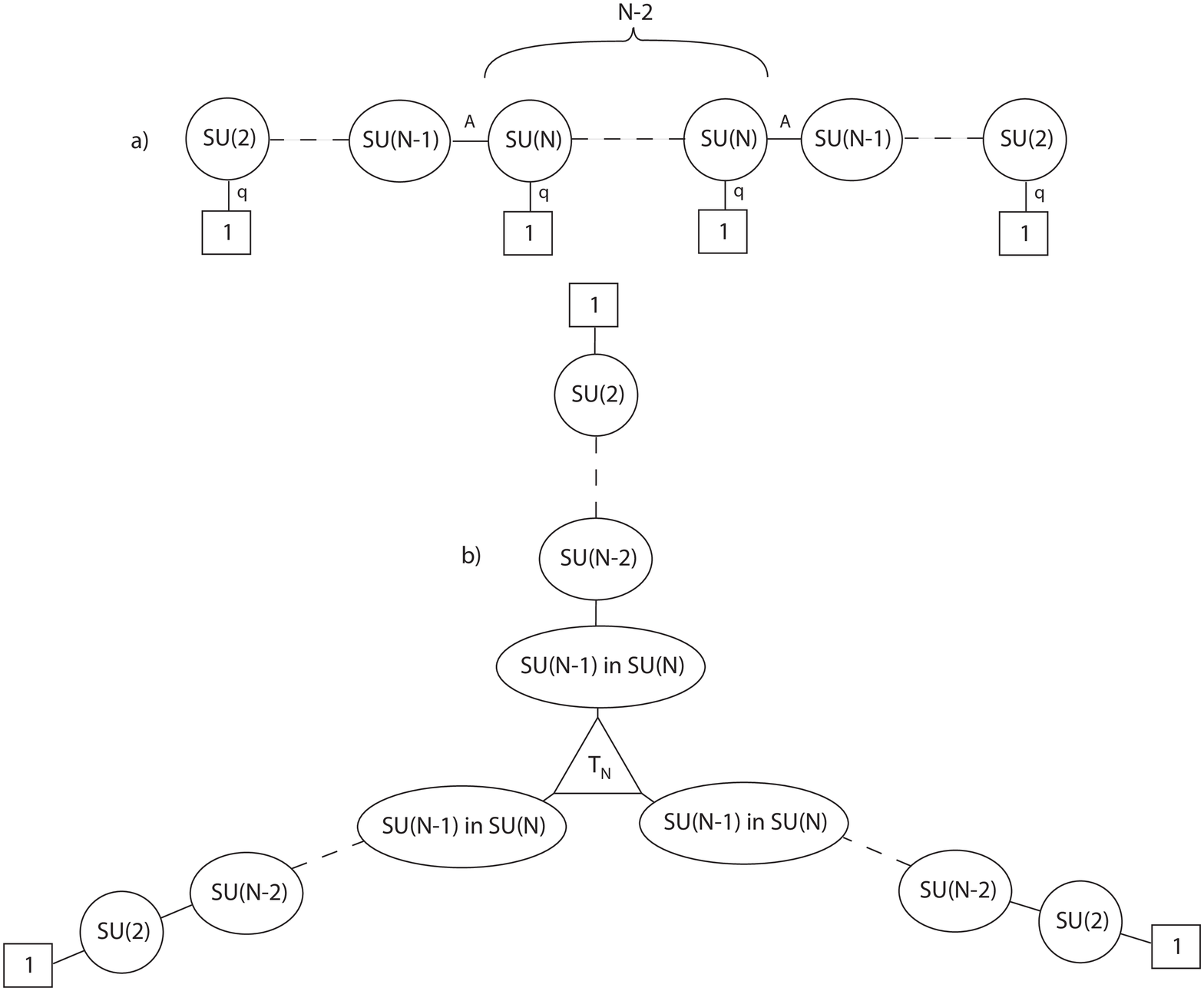} }
It is amusing to finally consider a symmetric setup, where a
sequence of nodes $SU(N-1)-SU(N-2) \cdots SU(2)$ is appended to all
three legs of $T_N$, see \symmetric (b).
 We can replace $T_N$ and any one of the three
sequences of nodes with a sequence of $N-2$ $SU(N)$ nodes, see \symmetric (a).
The resulting quiver gauge theory will be useful later in
the paper. It has only $3N-3$ $U(1)$ flavor symmetry groups, and is
associated to a sphere with $3N-3$ punctures, none special. It has a
single dimension $N-1$ operator, which is simply $H_{NN}$. The
uniqueness of the operator, and the $S_3$ symmetry of the setup
insure that it descends from a $O_{NNN}$ in $T_N$.

   \ifig\quivertn{ (a) Quiver constructed with the theory $T_N$ by gauging the $SU(N)$ global symmetries of
   the $T_N$ theory. Each circle indicates a gauge group. The resulting diagram can be viewed as giving a
   Riemann surface of genus two. We displayed two quiver constructions involving the $T_N$ theory that give rise
   to the same field theory. They corresponding groups are weakly coupled in different regions of the parameter space.
   In (b) we see a quiver associated  to  a higher genus Riemann surface. In this case it has genus four.
   Increasing the genus by one requires two extra $T_N$ factors as well as three extra gauge groups.
  } {\epsfxsize4.2in\epsfbox{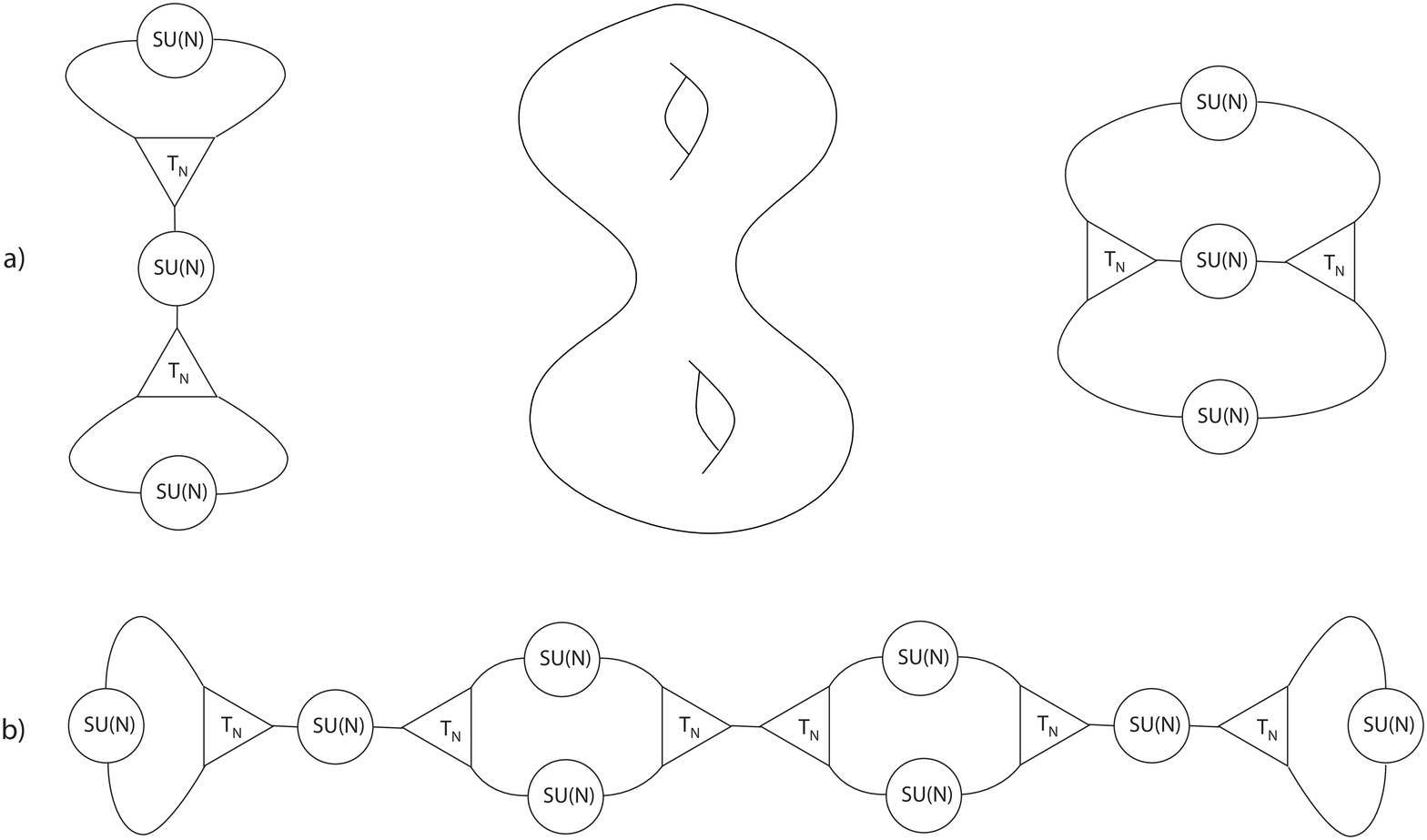} }

We can now use the theory $T_N$ as a building block for constructing
other theories. In particular, we can  build theories where all the
$SU(N)$ factors are gauged. We introduce a gauge group which is
gauging two of the $SU(N)$ global symmetries, belonging either to
the same $T_N$  or to two different $T_N$ theories. Such theories
are conformal because the $T_N$ contribution to the $SU(N)$  beta  function is
that of $N$ hypermultiplets.
 The  number of parameters of the resulting theory  is equal to the number of $SU(N)$ gauge fields
that we introduced. In fact,  if we link up all the $SU(N)$ factors, then we end up with a structure that
can be viewed as a higher genus Riemann surface with genus $g \geq 2$, see \quivertn .
Such a theory has $ 2 (g-1)$ $T_N$ building blocks and $ 3(g-1)$ $SU(N)$ gauge groups.

We can then easily compute the central charges by adding the contribution
of the $T_N$ theories and the vector multiplets. We can express the result in
terms of the effective number of vectors and hypers
\eqn\higherg{ \eqalign{
n^{g}_{v} = & (g-1) \left[ {4 N^3 \over 3 }   - { N\over 3 } -1  \right]
\cr
n^{g}_h = &  (g-1) \left[ { 4 N^3 \over 3 } - { 4  N  \over 3  } \right]
}}

From the operators $O_{ijk}$ of each $T_N$ theory we
 can construct gauge invariant operators by putting one
operator per $T_N$ factor and then contracting the indices in a gauge invariant fashion.
 This   gives
an operator ${\cal O} $ of  dimension \eqn\newdif{ \Delta = 2 ( g-1)
(N-1) }

\subsec{The geometry coming from $M5$ branes on a Riemann surface}

In this section we review the geometry constructed in \jmcn . It was obtained
by  a gravity construction which corresponds to starting
with the $M5 $ theory and putting
it on a  two dimensional Riemann surface and then flowing to the infrared.
In the UV we have a six dimensional
theory and in the IR we have a four dimensional theory. More concretely, one can start with a geometry
which is approximately $AdS_7 \times S^4$ at the boundary. The boundary geometry contains an
$R^4 \times \Sigma_2$. The $S^4$ is fibered over the $\Sigma_2$ in such away that we preserve eight supercharges.
When we flow to the IR we find that the geometry contains an $AdS_5$ factor and   six internal dimensions of
finite size.
These six internal dimensions can be viewed as an $S^4$ fibered over the Riemann surface. This
fibration is related to the twisting.
The Riemann surface we are considering has constant curvature and it is represented
as a quotient of hyperbolic
space.

More explicitly the full eleven dimensional geometry that
describes the IR fixed point is given by \eqn\metrieleven{
\eqalign{ { ds^2_{11}  } =&  ( \pi N l_p^3)^{2/3} { W^{1/3} \over 2}
\left\{   4 ds^2_{AdS_5} +   2 \left[ 4 { ( dr^2
+ r^2 d \beta^2 ) \over (1-r^2)^2 } \right]  + 2  d\theta^2  \right.+
\cr &\left. + { 2 \over W} \cos^2 \theta ( d\psi^2 + \sin^2 \psi d\phi^2  ) +
 { 4 \over W}  \sin^2 \theta ( d \chi +  { 2 r^2 d \beta \over (1-r^2) } )^2
\right\}
\cr
&~~~ W  \equiv  (1 + \cos^2 \theta )
}}
Here the angles $\theta, ~\psi, ~ \phi  , ~\chi$ parametrize a space which is topologically a four sphere.
The coordinates $r, \beta$ parametrize two dimensional hyperbolic space\foot{ The term between brackets [ ], including
the factor of four, is the metric on the unit radius hyperbolic space.}.
 In order to obtain a compact Riemann surface we are quotienting this hyperbolic space by
 a group $\Gamma$.
 The formula for the central charge is
$c = {\pi R^3_{AdS_5} \over 8 G_N^5}$, where $G_N^5$ is the five dimensional Newton constant  \HenningsonGX
 .
The eleven dimensional
Newton constant is $G_N^{11} = 16 \pi^7 l_p^9$. We then obtain that
\eqn\finre{
c =  { N^3 \over 3  }  {A_{\Sigma } \over 4 \pi }  =
{ N^3 \over 3 } (g-1) ~,~~~~~~~~A_{\Sigma } = 4 \pi ( g-1) ~,~~~~~~~~~g>1
}
where we used that for a Riemann surface with constant curvature the area is proportional to the
 Euler characteristic.
We see that this agrees with what we obtain from \higherg \ntwo , up to corrections of order $N$.
Note that the formulas in \higherg \ntwo\ also imply that $a=c$ up to corrections of order $N$.
Namely, $ 24 ( a-c ) =    (N-1) (g-1)$. The order $N$ term  arises from an $R^2$ correction in
$AdS_5$  \refs{\AharonyRZ,\BlauVZ,\NojiriMH}  . This, in turn, comes from
  the $R^4$ correction in eleven dimensions \GreenDI   .
 In fact, it can easily be seen that such a term has the correct $N$ dependence.
 The numerical coefficient can be checked as follows. First we note that we can also
 alternatively view $a-c$ as the anomaly coefficient involving the $U(1)_R$ current
 and two stress tensor insertions \ntwo . This four dimensional anomaly descends from the anomalies
 in six dimensions for the 5 brane theory. These anomalies were computed in \HarveyBX , by considering
 the gravity inflow contribution. As explained in \HarveyBX , that amounts to computing certain
 Chern Simons terms in seven dimensional gauge supergravity. Their final answer has a term
 involving purely the normal bundle  (purely the gauge fields of 7-d gauged supergravity) plus
 another term which is linear in $N$ whose total contribution is equal to $N$ times the contribution
 of a single fivebrane (see eqn (2.5) in \HarveyBX ). The contribution to $(a-c)$ must come from the
 term linear in $N$ since it is the only term that involves the tangent bundle.
 For a single fivebrane it is easy to compute $a-c$. A single fivebrane gives
 $g$ vector multiplets and one hypermultiplet. Thus a single fivebrane gives $24 (a-c) =  g-1  $,
 see \ntwo . Then $N$ fivebranes give  us the result that matches with the order $N$ term of the
 field theory answer. Note that this
 is a computation that was done entirely using the gravity dual, including the higher derivative corrections,
 since that is the method used in \HarveyBX .
 By a similar method it is possible also to match the subleading corrections to $a$ and $c$ individually.
 More explicitly one can consider the $U(1)_R SU(2)_R^2$ anomaly which is proportional to what
 we called $n_v$ \ntwo .  Then, according to the
 formulas in \HarveyBX , this anomaly is the sum of two terms. One is the contribution for a single
 M5 brane but multiplied by $N$. The second is a term that is proportional to $N^3-N$ (see eqn (2.5) in \HarveyBX ).
 This second
 term is the one we get from the leading order gravity result. On the other hand,
  the single fivebrane contribution has
 $n_v = g-1$. Note that the actual number of vector multiplets is $g$, but in this case the center
 of mass hypermultiplet fermion is charged under $SU(2)_R$ and gives the minus one contribution.
 (Ordinary field theory hypermultiplet fermions are uncharged under $SU(2)_R$.). Finally, the
 total contribution is simply the one we had from leading order supergravity with $N^3 \to N^3 -N$ plus
 $N$ times the one for a single brane. Then we have
 \eqn\finalans{
  n_v = { 4 \over 3 } (N^3 - N) (g-1) + N (g-1) = ( g-1) \left[ { 4 N^3 \over 3 } - { N \over 3 } \right]
  }
  This is the gravity result including the first subleading correction. It is almost equal to the field
  theory result \higherg . The extra (-1) we have in \higherg\ comes from the decoupled center of mass
  degree of freedom on the branes which is not included in the field theory. In the gravity side this
  could arise after quantizing all fields in this background, but we did not check it.
  It is similar to the computation done in \BilalPH , \AharonyDJ . The total form of $a-c$ is also
  proportional to $N-1$ and this minus one would arise in the same way from the center of mass
  degrees of freedom.

Let us now look for the state that is dual to the operator ${\cal O}$ described near \newdif .
This operator is an $M2$ brane wrapping the
Riemann surface $\Sigma$. The operator carries no $U(1)$ charge, so the M2 brane
  has to be located at a $U(1)$ invariant position. This means that $\theta =0$ in \metrieleven.
Using the formula for the tension of the $M2$ brane $T= { 1 \over ( 2 \pi )^2 l_p^3} $
and the area of
the Riemann surface at $\theta =0$ we find
\eqn\dimensic{
 \Delta_{grav}  =   { 1 \over ( 2 \pi)^2 l_p^3 } ( \pi N l_p ) 2     A_{\Sigma}
  = 2 (g-1) N
 }
 which agrees with \newdif\ up to a term of order one.
 This operator also carries $SU(2)_R$ spin $j =   (g-1)( N -1) $. The term proportional to $N$ arises
 as follows.  This M2 brane sits
 at point on the $S^2$. In addition, the M2 brane couples the four form flux $F_4$ which has a flux
 equal to $2 N (g-1)$ over $S^2 \times \Sigma_2$.
 This implies that the M2 brane behaves as
 a particle on $S^2$ with $2 N(g-1)$ units of magnetic flux over the $S^2$, which leads to
 $SU(2)_R$ spin $j = N(g-1)$.

 We expect that there is a correction to \dimensic\ that leads to
 a match with the field theory answer \newdif . We need a correction that adds
 $ -2 g + 2$ to \dimensic  . This correction should arise as a correction to the $SU(2)$ spin
 of the M2 brane and it  appears to arise as follows.
 The +2 is related to the quantization of the fermion  zero modes associated to the broken supersymmetries.
 These fermions are scalars on the M2 brane. We will later give an alternative argument for this fact.
 This M2 brane has additional bosonic and fermionic zero modes which are associated to one forms on
 the surface. The supersymmetric quantum mechanics will lead to a spin $j_R = g$ multiplet, as
 it was shown in a similar context in \GV . This can give rise to the $ -2 g $ correction, but we did not
 check the sign.

 It is actually possible to wrap an M2 brane multiple ($n$) times on
 the Riemann surface. If the M2 brane is connected, it takes the form of some genus
 $n(g-1)+1$ surface with covers the original one $n$ times. It is
 clearly possible to build the corresponding operator ${\cal O}_n$
 by taking $n$ copies of $O_{ijk}$ at each factor of $T_N$ and
 contracting indices to imitate the topology of the multiply wound
 M2 brane.

 Note that the group $\Gamma$ that produced the Riemann surface contains a number of parameters
 corresponding to the moduli of the Riemann surface. The number of parameters matches precisely
 the number of $SU(N)$ gauge groups which we used in the construction, namely $3 (g-1)$.
 Note that if we go near a degeneration region of the Riemann surface where it develops narrow
 tubes connected by various ``pants'', then we get a picture that corresponds closely
 to the field theory quiver in  \quivertn . Each tube corresponds to a gauge group, the length
 to width ratio of the tube is related to $1/g^2$. The amount of twisting of the tube as we join
 the two sides is related to the theta angle. If the tube is narrow, we expect a reduction to
 the metric of a D4 brane. Finally, the ``pants'' corresponds to the theory $T_N$.

 The reduction of the $C_3$ gauge potential in M theory on $\Sigma_2$ might lead to
 a U(1) gauge field, $A_1$,  in $AdS_5$ and a $U(1)$ global symmetry in the gauge theory.
 However in $AdS_5$ we would have a coupling $ N \int_{AdS_5} A_1 \wedge F_4 $, which is
 giving a mass to this gauge boson. Thus, we do not have an extra massless $U(1)$ gauge boson in $AdS$.
 Thus the theory does not have any $U(1)$ global symmetry.

 The theory should also have chiral primary fields which carry $U(1)$ charges coming from
 Kaluza Klein reduction. It would be nice to verify that we get the right spectrum of fields.
 This should match the spectrum of differentials of various degrees on the Riemann surface.
 Namely,  an operator  with  $U(1)_R$ charge $l$ in the six dimensional (0,2) theory
 becomes an $l$-th differential after the twisting. In other words, we can write the formal
 expression
  $Tr[ \phi_z^l ]$. We have that $l=2,3 \cdots, N$. It seems intuitively clear from the twisting
  procedure that the original KK modes of $AdS_7 \times S^4$ would lead to the correct
  spectrum of states, but we did not verify all the details.

This theory also has a very rich variety of Wilson and 't Hooft line operators. For each gauge
group in the quiver in \quivertn\ we have Wilson and a 't Hooft operator. In addition we have mixed operators.
These operators correspond to M2 branes with one direction wrapping a one cycle in the Riemann surface and
the other  two directions
along the $AdS_5$ directions, so that they intersect the $AdS_5$ boundary along a line.
They sit at $\theta = \pi/2$ which is an $SU(2)_R$ invariant point and they are localized on the $\chi$ circle.
The field theory operators are BPS operators including the gauge connection and the adjoint scalar.
It seems that the method proposed in \Pestun\ might lead to exact expressions for these observables.
Note that there are more gauge groups than there are topologically non-trivial cycles on the surface.
In fact, due to the presence of the operators $Q_{ijk}$ of the $T_{N}$,  not all the Wilson loops  are
topologically non-trivial. Of course, a Wilson loop does not need to be topologically non-trivial in
order to be BPS.

\newsec{ Punctures and more general field theories}

In the previous section we discussed configurations involving the IR limit of
$N$ M5 branes wrapping a
Riemann surface. In this section we add extra M5 branes that intersect this
Riemann surface at a points. The simplest possibility is to add one extra M5 brane
intersecting the original $N$ M5 branes at a point. We will first analyze this problem in
the probe approximation. We will later discuss what happens when we add a large
number of extra M5 fivebranes and discuss the full backreacted solutions near the
punctures. We also discuss a class of solutions with an extra $U(1)$ symmetry which
will display a precise match to the field theory discussion of the various punctures in \Gaiotto .

The most general eleven dimensional geometries preserving ${\cal N}=2$ superconformal symmetry
were found in \LLM .  They are given in terms
of  solutions of a Toda equation, as we review below. In this section we will explain in detail
the boundary conditions for the Toda equation which lead to well behaved geometries dual to the
field theories under consideration.

\subsec{ Probe approximation }

Let us first consider the simplest puncture.
This corresponds to adding a single M5 brane that is intersecting the Riemann surface at a point.
Of course, this M5 brane is also extended along the four dimensional spacetime.
This corresponds to adding a single probe M5 branes to the geometries we discussed in the previous
section. The M5 brane should also be extended along all of the $AdS_5$ directions and it
should preserve the $SU(2)_R \times U(1)_R$ symmetry of the superconformal algebra.
The requirement of $SU(2)_R$ symmetry implies that the brane should be sitting in the region where
the two sphere associated to the $SU(2)_R$ symmetry shrinks to zero. In addition, it should be wrapping
the circle associated to the $U(1)_R$ symmetry.
 Thus the brane
  sits at $\theta =\pi/2$ in \metrieleven ,  and at a point in the $H^2/\Gamma$ space.

   \ifig\probe{ (a) We add an extra gauge factor to   the
   configurations in \quivertn .  We have an extra bifundamental field $A$ .
   (b) This  gives rise to a Riemann
   surface with an elementary puncture. This adds  a single M5 brane that
   is transverse to the Riemann surface and wraps the circle corresponding to the $U(1)_R$ symmetry.
   It intersects the $N$ fivebranes that wrap the Riemann surface at a point on the Riemann surface.
  } {\epsfxsize2.7in\epsfbox{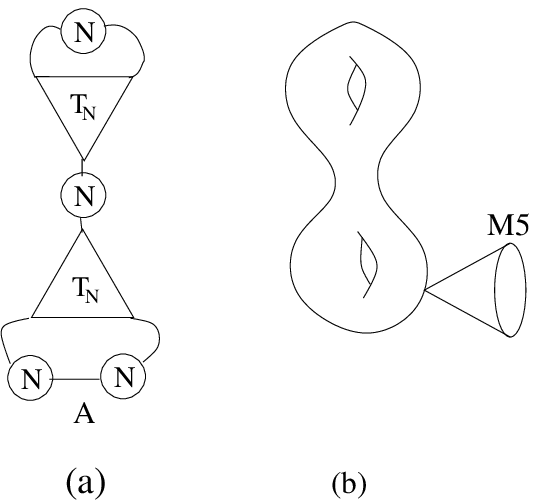} }

 In the field theory construction of the previous section,
 the addition of a single brane   corresponds to the addition of a single
 gauge group factor,   see \probe . This extra gauge group factor implies that
 now we have a bifundamental field $A$ , see \probe . This leads to an
 extra global $U(1)$ symmetry that acts on $A$ as $A \to e^{ i \varphi/N} A$. Note
 that $A \to e^{ i 2\pi/N} A$ is a gauge transformation, so that $\varphi \sim \varphi + 2 \pi$.
 With these bifundamentals we can construct the
 BPS operator   $Det(A)$ which has charge one under this  global
 $U(1)$ symmetry.  In addition,  it has $SU(2)_R$ spin  $j=N/2$ and   conformal
 dimension $N$.

 Returning to the gravity description, we note that the extra
  probe brane has a worldvolume given by $AdS_5 \times S^1$.
 The M5 brane has a two form potential with a self dual three form field
 strength on its worldvolume.
  The Kaluza Klein reduction of this two form
  potential on the $S^1$ gives rise to a $U(1)$ gauge field in $AdS_5$. This
  corresponds to the global $U(1)$ symmetry rotating the phase of the bifundamental
  field $A$. We also have a BPS state given by
    an M2 brane which ends on the fivebrane along the
 $S^1$ and it is extended from $\theta = \pi/2$, where the
 $M5$ sits to $ \theta =0$ where the $S^1$ shrinks. This brane has charge
 one under the gauge field in $AdS_5$ associated to the global $U(1)$ symmetry.
  One can easily check from \metrieleven\ that such a brane
  has dimension $N$. Its $SU(2)_R$ spin arises from the flux of $G_4$ on
  a fourcycle given by the two sphere together with the
   two cycle wrapped by the  M2 brane.
  This fourcycle is, topologically, the $S^4$ transverse to the $N$ original M5 branes wrapping
  the Riemann surface.  Thus the flux of $G_4$ is $N$ and the $SU(2)_R$ spin is $j=N/2$.

 \subsec{ A general class of gravity solutions which are
  dual to  ${\cal N}=2$ superconformal field theories}

 The most general gravity solution of eleven dimensional supergravity which preserves ${\cal N}=2$
 superconformal symmetry was constructed in \LLM \foot{ This was shown up to a small caveat discussed in
 appendix F.8 of \LLM . A particular flux was assumed to be zero and it was checked that it could not be turned
 on infinitesimally, but it is possible that there are solutions where such a flux is finite and cannot be turned
 off infinitesimally. It would be nice to close this possible loophole, or find a more general ansatz, if that were
 possible.}.
  The full solution is constructed from a single function $D(x_1,x_2,y)$
  which depends on three variables and
  it obeys a Toda equation of the form
  \eqn\toda{
  \partial_{x_1}^2 D + \partial_{x_2}^2 D  + \partial_y^2 e^D =0
  }

  A simple solution to this equation is
  \eqn\solsymp{
  e^D =   ( N^2 -y^2) { 4 \over ( 1- r^2 )^2   }  ~,~~~~~~~~r^2 = x_1^2 + x_2^2
  }

  When this solution is inserted  into the ansatz in \LLM\ we recover \metrieleven .
  For the reader's  convenience we reproduce the  ansatz
  \eqn\ansatz{ \eqalign{
  ds_{11}^2&= \kappa^{2/3} e^{2  \tilde \lambda}\left( 4
ds_{AdS_5}^2+y^2e^{-6{\tilde\lambda}} d{\tilde\Omega}_2^2+
ds_4^2\right)  \cr
ds_4^2&= { 4  \over (1 - y \partial_y D )  } (d\chi+v_idx^i)^2 +
 { - \partial_y D \over y }  [ dy^2 + e^{D}
(dx_1^2 + dx_2^2) ]
\cr
v_i &= \half \epsilon_{ij} \partial_j D ~~~~~~~ v= v_i dx_i
 \cr
 e^{-6\tilde \lambda}&= - {\partial_y D \over y(1 -  y \partial_y D) }
 \cr G_4 = & \kappa F_2 \wedge d\Omega_2
 \cr
 F_2 =  &  2 \left[ ( dt + v) d( y^3 e^{- 6 \tilde \lambda } ) + y ( 1 - y^2 e^{ - 6 \tilde \lambda} )
 d v - { 1 \over 2 } \partial _y e^D dx_1 dx_2 \right]
 }}
 This is the most general M-theory solution which preserves four dimensional
 ${\cal N}=2$ superconformal symmetry. Here we introduced a constant $\kappa$ multiplying the metric
 which we could remove by performing a rescaling of $y$ and $e^D$. We introduce it to simplify
 the charge quantization conditions.
 We will mostly take $\kappa = { \pi \over 2 } \,  l_p^3 $, but occasionally, in
 some solutions we will set it to other values\foot{ In these conventions
 $ { 1 \over (2 \pi)^3 l_p^3 } \int G_4 $ is an integer, up to the effects discussed in \wittenquantization .
 The coupling of $C_3$ to the M2 is via $\exp\{ { i \over ( 2 \pi)^2 l_p^3 } \int C_3 \}$. }.

 The region $y=0$ is special because   $S^2$ shrinks. This is ensured
 by demanding that $e^D$ is finite and $\partial_y D =0$ at $y=0$. This  condition
 is  obeyed in the regions with no punctures. Below we will derive the correct condition when
 we have a puncture.
The circle parametrized by $\chi$
 shrinks at $y=y_c$. This happens in a non-singular fashion if
  $e^D \sim (y_c-y)$ near $y\sim y_c$.
 Notice that in this region $\partial_y D \to \infty $ and $e^{  3 \lambda } = y_c$.
 The $\chi$ circle is   associated to the $U(1)_R$ symmetry.
  More precisely, a $U(1)_R$ transformation shifts the coordinate $\chi$.

  We can  construct an $S^4$ by taking the interval
  from $y=0$ to $y=y_c$ together with the $S^2$ and the $\chi$ circle. We can easily compute
  the flux of the four from through this sphere and   we find that the flux is proportional
  to $y_c$. Choosing $\kappa = { \pi \over 2} \, l_p^3$,    the flux quantization condition leads to
    $y_c = N$. Thus, the value of $y_c=N$ is constant over the Riemann surface, and
  we identify it with the number of wrapped branes. This is true even in solutions where we add an arbitrary
  number of punctures, including backreaction.

 \ifig\fluxes{ (a) Various four cycles we can use to measure fluxes. If we consider the line from $y=0$ to $y=N$
 together with the $S^1_\chi$ and the $S^2$ we form a four sphere. The flux on this four sphere is equal to $N$
 and it is counting the number of M5 branes wrapping  the Riemann surface. When we add a probe brane at $y=0$, we can
 form a four cycle from the ``cup'' in the $y,x_1,x_2$ coordinates displayed in (a) together with the $S^2$.
 The flux over this four cycle counts the number of probe M5 branes. In (b) we show a BPS state which comes
 from an M2 wrapping the interval together with the $S^1_\chi$.
 } {\epsfxsize4in\epsfbox{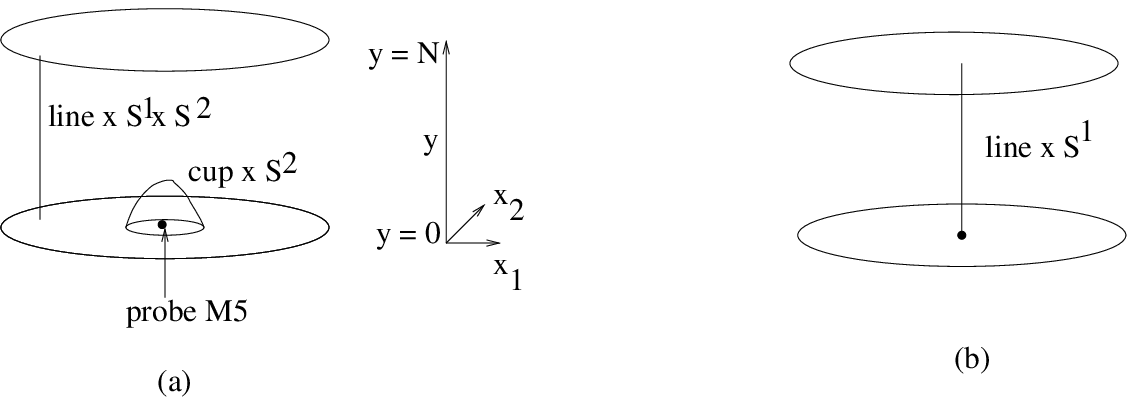} }

 We can add a probe M5 brane at $y=0$ which is wrapping
 $AdS_5$ and the circle. Such a probe sits at a point in the $x_1,x_2$ coordinates.
 This is basically the probe that  we discussed in the previous subsection, except that now
 we are adding it to an general solution.
 One can surround the probe brane with an $S^4$ which measures its flux.
 This $S^4$ consists of the $S^2$ of the $SU(2)_R$ symmetry and a small cup in the $y,x_1,x_2$ coordinates
 that surrounds the probe brane, see \fluxes (a).
  This surface is topologically an $S^4$ since the $S^2$ shrinks at
 the rim of the ``cup''. The flux on this $S^4$ counts the number of probe M5 branes.

  We can now consider the BPS state given by
  an $M2$ brane wrapping the $\chi$ circle and ending  on the probe M5 at $y=0$.
  The brane extends from $y=0$ to $y=N$ where the circle shrinks, see \fluxes (b).
  We can easily compute the mass from   \ansatz\ and we find that it is equal to $N$.
 In addition,  the flux of the four form field strength on
 this surface times the  $S^2$ is also  $N$, which gives the M2 an $SU(2)_R$ spin $j=N/2$.

\subsec{ Full back reacted solution near the puncture}

In this section we   analyze in more detail the gravity
solution near a probe brane and we  find
the full backreacted solution.
 For this purpose, let us first imagine we have a  background like \solsymp\
 with a very large number of branes $N$.
In addition, we now add a number $K$ of coincident  M5 brane probes, as we discussed in
the previous subsection.
If $K$ is sufficiently large, but $K \ll N$, then we can analyze the
 problem locally near this puncture.
Near this region, we  have a system of $K$ M5
branes wrapping $AdS_5 \times S^1$. Its near horizon
geometry is very easy to find since $AdS_5 \times S^1$ is conformal
 to $R^{1,5}$. Thus the near horizon geometry
is simply a rewriting of the $AdS_7 \times S^4$ geometry.
In fact we can write $AdS_7$ as
\eqn\adsevrew{
 ds^2 =   \cosh^2 \zeta  ds^2_{AdS_5} + \sinh^2 \zeta d\varphi^2 + d\zeta^2
 }
 where $AdS_5$ has radius one and $\varphi $ has period $ 2 \pi$.
 For large $\zeta$ the boundary is $AdS_5 \times S^1$. As we go to the interior we see that
 the $S^1$ shrinks in a smooth way and the $AdS_5$ warp factor remains bounded below. This is a smooth
 horizon-free metric. In conclusion, we expect that the metric
 close to a probe brane looks as in \adsevrew .

 \ifig\rods{In (a) we schematically display the solution for $AdS_7\times S^4$. There is a line
 source that starts at $y=0$ and extends to $y=2 K$. In (b) we display a ``cup'' surrounds the rods. Together with
 the two sphere this forms a fourcycle. The flux on this fourcycle is $K$, the number of transverse fivebranes.
  In (c) we display the problem that would
 correspond to the insertion of many punctures. We have line charge sources for the Toda equation
 on several segments located at different positions on the Riemann surface parametrized by
  $x_1,x_2$.
 } {\epsfxsize4in\epsfbox{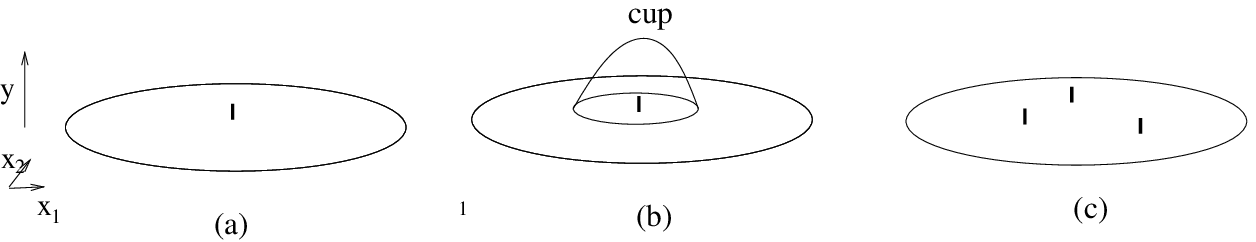} }

 We now rewrite this metric, \adsevrew , in terms of the general ansatz \ansatz .
 We do this in order to understand the
 type of boundary conditions for the Toda
  equation which locally lead to a non-singular geometry such as
 \adsevrew .

The starting point is $AdS_7 \times S^4$ written as
 \eqn\startad{
  ds^2 = 4
(  \cosh^2 \zeta ds^2_{AdS_5} + \sinh^2 \zeta d \varphi^2 + d\zeta^2
) + d \theta^2 + \sin^2 \theta d\psi^2 + \cos^2 \theta ds^2_{S^2}
} We define the angles \eqn\angldef{ \varphi = \chi
~,~~~~~~~~\psi = 2 \chi + \beta
   }
This implies that \startad\ becomes
 \eqn\midad{
 \eqalign{ ds^2 = &
4 \cosh^2 \zeta ds^2_{AdS_5}  + \cos^2 \theta ds^2_{S^2} + 4 (
\sinh^2 \zeta + \sin^2 \theta) \left[
 d\chi + {1 \over 2 } {   \sin^2 \theta \over \sinh^2 \zeta
+ \sin^2 \theta } d \beta \right]^2  + \cr & + 4 d \zeta^2 + d
\theta^2 + { \sinh^2 \zeta \sin^2 \theta \over \sinh^2 \zeta +
\sin^2 \theta } d    \beta ^2
 }}
We then find $ e^{ 2 \tilde \lambda} = \cosh^2 \zeta $ and the
relations \foot{Here we are setting $\kappa =1$ in \ansatz .}
 \eqn\findres{
 y = \cosh^2 \zeta \cos \theta ~,~~~~~~
 e^{D} = { \cosh^2 \zeta \over \sinh^2 \zeta} ~,~~~~~~r = \sinh^2
 \zeta \sin \theta
 }
 where $dx_1^2 + dx_2^2 = dr^2 + r^2 d
\beta^2 $.

We are finding that $e^D$  goes to zero along a segment that sits at $r=0$ and
extends from $y=0$ to $y=1$. Near this segment we have $e^D \sim { \sqrt{ 1 - y^2} \over r } $.
Thus we have a constant charge density along this segment. The Toda equation is
getting a source on the right hand side of the from
\eqn\todanew{
\nabla^2 D + \partial_y^2 e^D = - 2 \pi  \delta^2 (r) \theta ( 1-y )
}
where  $\theta (1-y)$ is a step
function which is zero for $y>1$ and one  for $ 0 \leq y \leq 1 $. This a constant line
charge density in the three dimensional space parametrized by $y, ~x_1,~  x_2 $.
This is a singular solution of the Toda equation. However, the full ten dimensional geometry
is non-singular.

Reinstating our previous choice for $\kappa$ in \ansatz , and quantizing the flux appropriately,
we find that in  the case with $K$ fivebranes  the line charge density is still
of one, but  it extends to $y= 2 K$. Thus increasing $K$ leads to a longer rod.

We have learnt that we can solve the Toda equation adding charge sources that extend from $y=0$ to
$y=2 K_i$, and sit at positions $x_i$. The elementary puncture has charge $K=1$ and corresponds to a
very small rod. Even though the geometry is formally smooth, we cannot trust the solution near
a rod with $K=1$ because the curvature is of the order of the Planck scale.
One can compute the flux of $F_4$ around these rods, and we get that it is $K_i$.
This a flux computed on a surface that consists of the $S^2$ and cup surrounding the rod, see \rods (b).
In the next section we will show an easy way to do this computation.

Thus, a general solution on a Riemann surface with a number of punctures can be found by solving
the Toda equation with these boundary conditions.
 We can  start with a negatively curved Riemann surface  with no punctures and then start adding
  punctures. The
 solution will continue to look
qualitatively as we described so far.

\ifig\fluxcups{In (a) we display a solution with many punctures. We start with a fourcycle
at $y=N$ consisting of the Riemann surface $\tilde \Sigma $ times the $S^2$.
 In (b) we have moved this cycle down to an intermediate value of $y$. Due to the fact that the
 $S^1_\chi$ circle is non-trivially fibered we pick up a piece which is winding along the $\chi$ circle
 and extended in $y$.
  In (c) we display the final form for the fourcycle. We have a series of ``cups'' which surround
  the punctures and we have $2(g-1)$ times the $S^4$ which is transverse to the $N$ branes.
 } {\epsfxsize4in\epsfbox{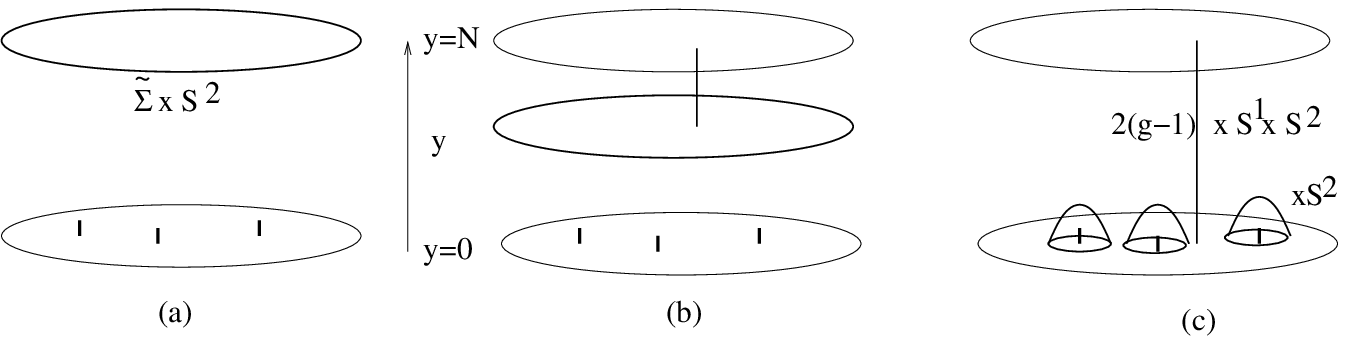} }

Let us make a few remarks about the solutions with a general number of punctures which do not
require finding the explicit metric.

When we introduce punctures there are  changes to the dimensions of some operators.
For example, let us start with the operator ${\cal O}$ that we discussed in \newdif\ and   add
a puncture. In order to construct a gauge invariant operator we now have to add the link
field $A$ in \probe . By a similar reasoning,  if we add  $K$ punctures
  the dimension of ${\cal O}$
increases by $K$ units. The operator ${\cal O}$ sits at $y=N$, and naively does not seem to care
about punctures. However, the addition of punctures changes the area of the two cycle $\tilde \Sigma $
that
the brane is wrapping. Note that at $y=N$, where the circle shrinks and $e^D \to 0$,
we have  $e^{ 3 \tilde \lambda } = N$. Then the  energy of an M2 brane wrapping the two cycle $\tilde \Sigma $
 is given by  the integral of $\partial_y e^D dx_1 dx_2$. We can compute it by noticing that the same
 integral gives us the flux of $G_4$ on a four cycle that consists of $\tilde \Sigma \times S^2$.
 The relation between the area and the flux is not surprising since the flux will induce an $SU(2)_R$ spin
 for this brane and the BPS condition links this spin to the energy of the brane. This fourcycle sits at $y=N$.
We can now deform it  and take it to $y=0$. This   picks up the contribution of
all the punctures, since we will be integrating on cups around the line sources, see \fluxcups (c).
  In addition there is an extra contribution which is due to the fact that the
circle $S^1$, parametrized by $\chi$,
 is non-trivially fibered over the Riemann surface. The $U(1)$ field strength
of this fibration is given by $dv = { 1 \over 2 } ( \partial_1^2 + \partial_2^2 )  D =
{ 1 \over 2 } \sqrt{ g^{(2)}} R^{(2)}  $. When this is integrated over the Riemann
surface we get $ \int dv = 4 \pi (g-1) $.   Thus when we deform the four cycle from $y=N$ to $y=0$
 we  end up
wrapping the $\chi $ circle $2 (g-1)$ times. We then conclude that the   energy of
 the wrapped brane, which is proportional to the area or the flux, is
\eqn\totalv{
\Delta  = { 2 \kappa \over (2 \pi)^2 l_p^3 }  \int_{ \tilde \Sigma }  ( -  \partial_y e^D )
dx_1 dx_2 = 2 (g-1) N + \sum_i K_i
}
This formula is valid for all $g$, including $g=0,1$. In addition,  the quantum mechanics on the brane is expected
to add a term of the form $2 - 2 g$ to this formula, giving
\eqn\fulldel{
 \Delta = 2 (g-1) (N-1) + \sum_i K_i
 }

Let us discuss in more detail the case of the sphere, $g=0$.
Note that \totalv\ is positive only if   $K_T = \sum_i K_i > 2 N $. Thus the classical geometry we
are proposing only makes sense if this inequality is obeyed. If the total number of punctures is large,
we can distribute them more or less uniformly on the sphere.
We can then  write an approximate
 solution of the form
 \eqn\approxso{
  e^D  \sim {  ( K_T - N -y) (N-y) { 4 \over ( 1 + r^2 )^2 } }  ~,~~~~~~~~~~~K_T = \sum_i K_i
  }
  where we fixed the normalization constant by the above considerations \totalv .  This solution breaks down
  at small $y$. At small $y$ we would need consider a solution of Toda with sources at specific positions.
  It is interesting to note what happens as $K_T = 2N$. In that case, we see from \totalv\ that the area
  of the two sphere becomes zero. This is {\it not} the two sphere associated to the $SU(2)_R$ symmetry
  but the two sphere of the Riemann surface that the branes are wrapping.
   In fact, we can check that if $e^D$ has a double zero of this kind, then
  the geometry near this region contains an $A_1$ singularity. The directions transverse to the singularity
  are given the $\chi$ direction, the $y$ direction and the worldvolume $S^2$ that is shrinking.
  On the $A_1$ singularity we have an $SU(2)$ symmetry. The operator \fulldel\ in this case has dimension
  $\Delta =2$ and it is the partner of the $SU(2)$ global currents.
   The case with $K_T=0$ was studied in \jmcn , by considering the gravity solution corresponding to
  an M5 brane wrapping the two sphere. It was found that the solution is singular in the IR. Thus, this
  theory appears not to flow to a conformal field theory in the IR. It would be nice to know what happens
  for $0<K_T < 2N$.

 Let us now consider the torus, $g=1$. From the field theory side, we know that
  when we add
 zero or one puncture, we expect to get $N=4$ super Yang Mills. The solution with $k$ punctures
 corresponds to the quiver that results from putting $N$ D3 branes at an $A_{k-1}$ singularity \mdgm.
 The corresponding gravity solution is $AdS_5 \times S^5/Z_k$ \skes . The field theory is a quiver diagram
 with the topology of a ring.
 We can write the $S^5$ as
 \eqn\fivesphe{
 ds^2_{S^5} = d\alpha^2 + \sin^2 \alpha d \chi^2 + { \cos^2 \alpha  \over 4 } \left[ ( d \psi + \cos \theta d \varphi)^2
 + d\theta^2 + \sin^2 \theta d\varphi^2 \right]
 }
 The $Z_k$ quotient takes $\psi \to \psi + { 4 \pi \over k }$.  We can now T-dualize $\psi$ and then lift to
 M-theory\foot{   H. Lin has worked   this out  in detail a few years ago.
 We thank him for discussions on this. }. The solution lifts up
 to
 \eqn\thorussol{
 e^D = { k \over 2 \kappa \tau_2  } ( N -y )
 }
 where $\tau_2$ is the $\tau$ parameter original IIB coupling constant.
 For large $k$ and large $N$ the M-theory description becomes good. It is given by \thorussol\ as long
 as $y$ is large, but as $y\to 0$ we should replace it by a solution which depends more explicitly on where
 the punctures are.

Note that the field theory discussion in \refs{\wittenM,\Gaiotto} involved a Riemann surface with punctures.
More precisely, it was argued that the space of coupling constants is the same as the moduli space of
a Riemann surface with punctures. For that purpose only the complex structure of the space was important.
In these gravity solutions it is natural to identify that Riemann surface with the surface parametrized by
$x_1,x_2$ at $y=0$. The metric on this surface comes from solving the Toda equation with the appropriate
boundary conditions. Another natural Riemann surface is the surface $\tilde \Sigma$ which is wrapped by
the BPS M2 brane
that  sits at $y=N$. This surface has no punctures.

Note that one could consider the six dimensional theory corresponding to M5 branes wrapping a Riemann surface.
In that solution one is allowed to fix any metric that one wishes on the surface. However, we expect that
when we flow to the IR fixed point, which is a four dimensional field theory, then all the detailed metric information
on the surface is washed out and the IR fixed point depends only on the complex structure. It would be nice
to check this explicitly by finding the gravity solution describing a flow with arbitrary metrics.
The flows discussed in \jmcn\ assumed a constant curvature metric for the surface also for the six dimensional
field theory.

Conversely, starting from the quiver theories described here it is possible to take a limit
where one would recover the M5 brane on a Riemann surface with an arbitrary metric.
One can due this by a simple generalization of the deconstruction idea in \deconstruction .
Let us start with a quiver gauge theory with a large number of punctures, $K\gg N$.
 Let us now consider the same Riemann surface in the geometry at $y=N$ where we wrapped the M2 brane which
 lead to \totalv . Now we wrap an M5 brane instead. It is an M5 brane which sits at some value of the
 radial position of $AdS_5$. This brane corresponds to the Higgs branch of the theory. We are breaking the
 $SU(2)_R$ symmetry. Note that the formula \totalv\ implies that the area of the Riemann surface in
 units of the radius of $AdS_5$ is  proportional to  $  A_2 \sim K/N + 2 (g-1) \sim K/N $. This implies that
 when we take the limit $K/N \to \infty$ we will get a very large surface. In fact, when  we put
 all $N$ branes into the Higgs branch, on top of each other,
  and we take the $K\to \infty$ limit that limit we reconstruct
 the six dimensional
 (0,2) theory of $N$ M5 branes on the Riemann surface. In order to define that theory we need to specify the
 scale factor of the metric on the surface. This is simply given by the local density of punctures as we take
 the limit. Thus if we take the punctures uniformly distributed, we get a constant curvature metric, while if
 we take another distribution we will get a different metric on the two dimensional surface that differs by
  a Weyl factor. The number of six dimensional degrees of freedom can be read off by considering the
  four dimensional number, which is $N^2 K $,  and writing it as $N^3 { K \over N} = N^3 A_2$.
  The case considered in \deconstruction\ is the special case of the torus.

\subsec{ Detailed study of the possible punctures }

In this section we study the various types of punctures we can have. The simplest is the
one corresponding to a single M5 brane. The others arise when we take many M5 branes together and take various
limits. A rich set of punctures was found in \Gaiotto .
We describe explicitly the corresponding solutions.

We assume that we have a configuration which is $U(1)$ symmetric around the puncture. So we look for
solutions that have a rotational symmetry in the $x_1,x_2$ coordinates of \ansatz .
In that case, we can  use the equivalence between the $U(1)$ symmetric solutions of Toda and
axially symmetric electrostatics problems in three dimensions \ward .
  \eqn\wardtrick{
 {r^2  e^{D}  }  = \rho^2 ~,~~~~~~~~~~y=\rho \partial_\rho V \equiv \dot V ~,~~~~~~~~  \log r =\partial_\eta V \equiv V'
 }
 Then the Toda equation \toda\ becomes the cylindrically symmetric Laplace
 equation in three dimensions
 \eqn\electros{
 {1 \over \rho } \partial_\rho ( \rho \partial_\rho V) + \partial^2_\eta V =  0 ~~~~{\rm or}~~~
  \ddot V + \rho^2  V'' =0
 }

 Now the full eleven dimensional geometry can be specified in terms of the function $V$.
 It is a simple matter to write the ansatz in \ansatz\ in terms of $V$   \hljm
  \eqn\soltuc{\eqalign{
  d s_{11}^2 =& \kappa^{2/3} \left( {   \dot V \tilde \Delta \over 2 V'' } \right)^{1/3}
   \left[ 4 ds^2_{AdS_5} +   { 2  V'' \dot V \over \tilde \Delta} ds^2_{S^2} +
   {2  V''\over \dot V} ( d \rho^2 + {   2 \dot V \over 2 \dot V - \ddot V }\rho^2  d\chi^2 + d\eta^2 )
   + \right.
   \cr
    & ~~~ + \left. { 2 (  2 \dot V - \ddot V  ) \over \dot V \tilde \Delta }
     ( d \beta + { 2 \dot V \dot V' \over 2 \dot V - \ddot V} d \chi )^2 \right]
\cr
\tilde \Delta = & (   2 \dot V - \ddot V ) V'' + ( \dot V')^2
\cr
C_3 = &  \kappa 2  \left[ - 2 { \dot V^2 V'' \over \tilde \Delta } d\chi +  ( { \dot V \dot V ' \over \tilde \Delta } - \eta ) d\beta
\right] d\Omega_2
 }}
where both $\chi$ and $\beta$ have period $2 \pi$.

  \ifig\linecharges{ In (a) we see the type of electrostatic problem we need to solve. It is a three dimensional
  problem with axial symmetry around the  axis $\eta$. There is conducting disk at $\eta=0$ and
  a line charge density at $\rho =0$.  In (b) we see the line charge density as a function of $\eta$ for the
  hyperbolic disk solution. In (c) we see the charge density for the $AdS_7 \times S^4$ solution. In
  (d) we see the line charge density for a certain number of branes at the origin of the hyperbolic disks. It
  results from putting (b) and (c) together.
 } {\epsfxsize4in\epsfbox{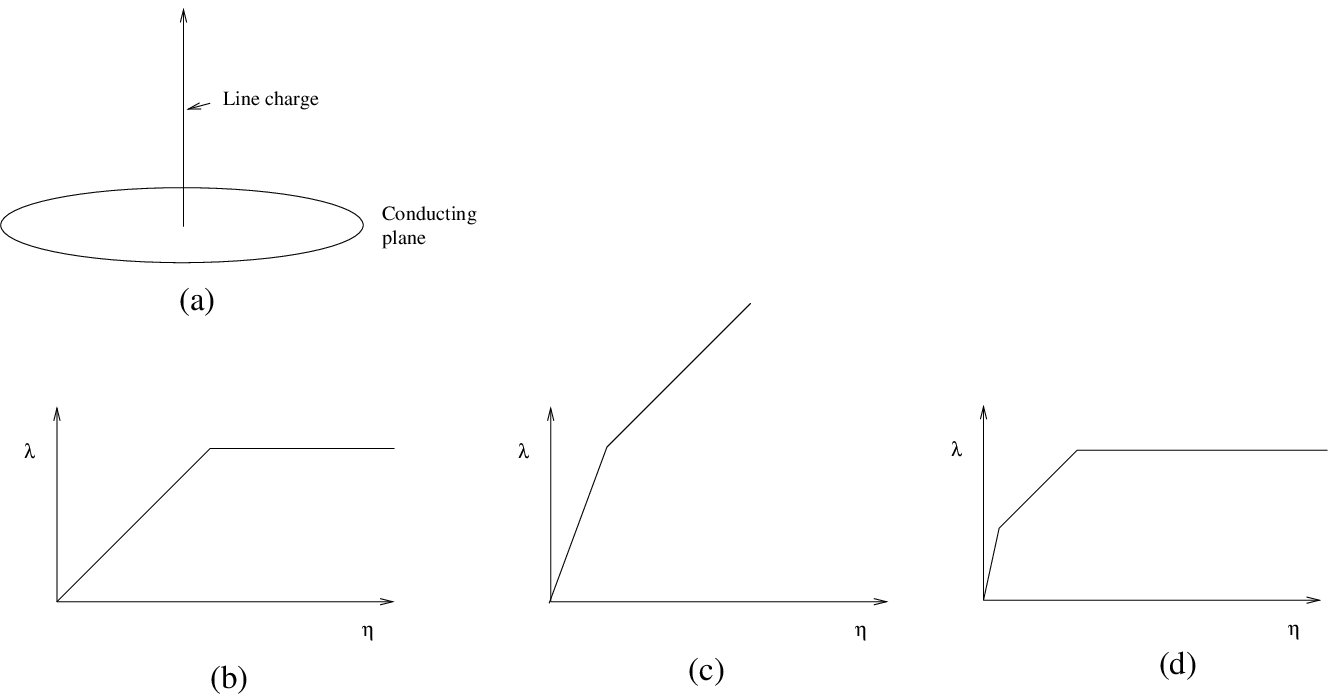} }

   \ifig\profiles{ We see a couple of quivers and their corresponding charge densities. The quivers are ``tails''
   which are ending a quiver with ${\cal N}=2$ supersymmetry. The eta values $\eta =1,2,3$ denote the first, second,
   third gauge groups. The values of the line charge density $\lambda(\eta)$ at integer values of $\eta$
   correspond to the ranks of the $SU(N)$ gauge groups at the corresponding nodes. Each time the slope changes by
   $k$ units, there are $k$ extra fundamentals.
 } {\epsfxsize4in\epsfbox{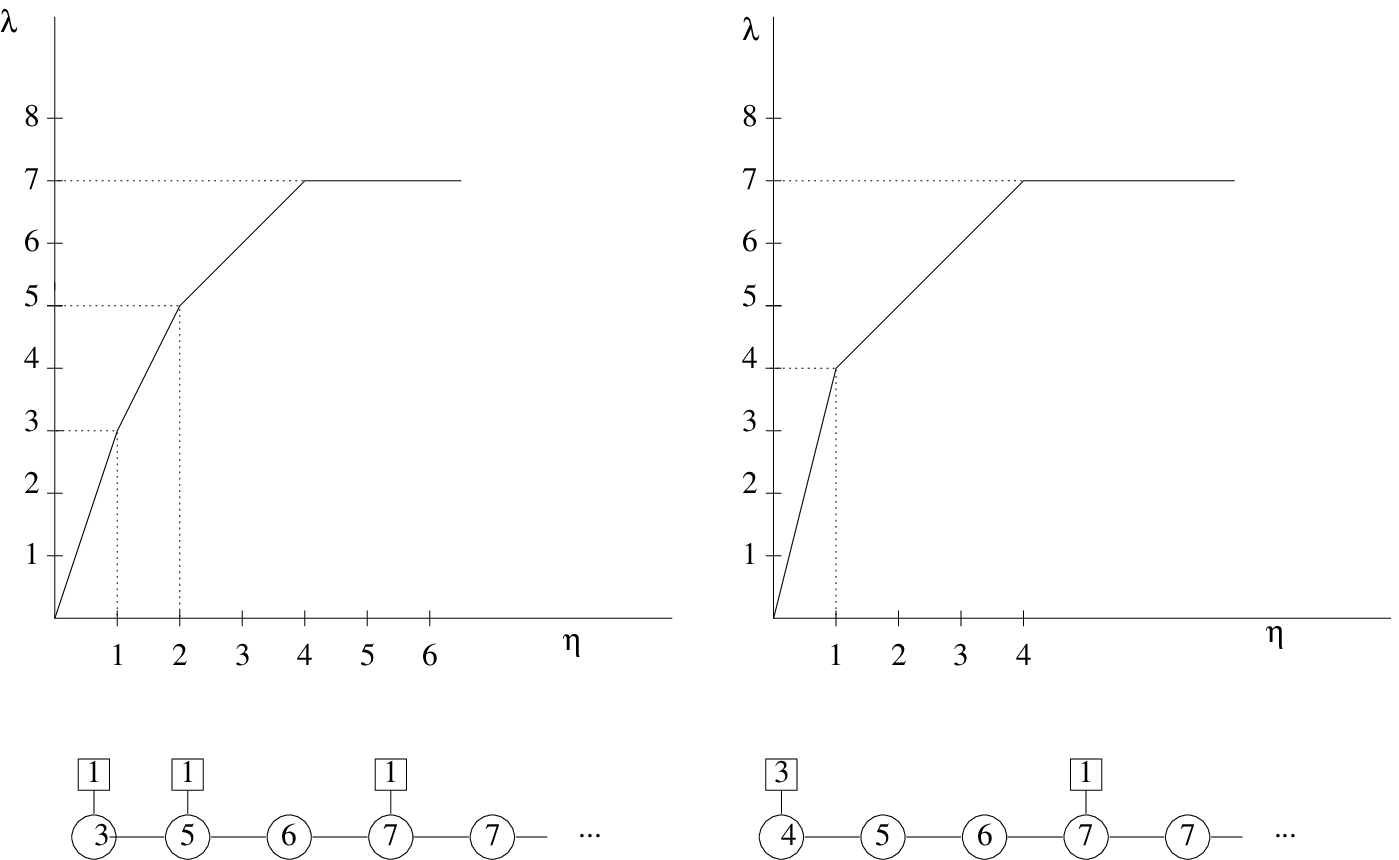} }

 We need to understand what electrostatic problems lead to reasonable geometries.
  By considering some examples it is possible to guess the general rule.

   The first example is the solution in \findres . We can compute the variable $\eta$ using \wardtrick\ and
   we find $\eta = { 1 \over 2 } \cos \theta \cosh 2 \zeta $.
    We then see from \findres\ that
   $\rho =0$ arises only when $r=0$.  Note that since $y$ is generically nonzero when $\rho=0$ this means from
   \wardtrick\ that there is
   a line charge density $\lambda(\eta) = y(\eta, \rho=0)$ along the  $\rho=0$ axis.
    We can have
    $r=0$ either because  $\theta =0 $  or because  $\zeta=0$ in \findres .
   In the first case we find that $y = \lambda(\eta) =  2 \eta$, with $ 0 \leq \eta \leq { 1 \over  2 } $.
   In the second case that $y = \lambda(\eta) = \eta + { 1 \over 2 } $, $\eta \geq { 1 \over 2 } $. In the end
   we have a line charge density as depicted in \linecharges c . In addition we have a
   conducting plane at $\eta =0$. So we only consider the region $\eta \geq 0 $. The conducting
   plane corresponds to the region where the $S^2 \to 0$ at $y=0$.

   The second example is simply the solution \solsymp .
   We have
   \eqn\translat{
   r^2 e^D = \rho^2 ~,~~~~~~~~~ \eta =   y { (1 + r^2 ) \over (1 - r^2) }
   }
   We now see that when $\rho =0$ we can either have $r=0$ or $y=N$. In the first
   case  we get a linear  charge density with unit
   slope $ y =    \lambda(\eta ) = \eta $, $  0 \leq \eta \leq N $.
    Or we can have $y=N$ and then we see that $ \lambda = y =N$ is a constant, and $  N \leq \eta $.
    The final picture for the charge density looks as in \linecharges (a) .

    In both of these examples we can rescale the $y$ coordinate, together with $e^D$.   This is
    a symmetry of \ansatz\ , which corresponds to rescaling the metric.

    Finally, let us discuss the third example. This is simply a $Z_k$ quotient of
    the hyperbolic space solution. We simply set $\beta \to \beta + { 2 \pi \over k }$.
    We can now define a rescaled $\tilde \beta = k \beta$ and then put the full metric
    in terms of the
    $U(1)$ symmetric ansatz. One has to follow the effects of this rescaling.
     It turns out that the effect
    of the rescaling is simply to modify the
     slope of the line charge between $\eta =0$ and $\eta=N/k$ to
    $\lambda( \eta) = k \eta $.
    The metric has an $A_{k-1}$ singularity which is located
    at $r=0$, $y=N$ in the original coordinates.
    In the $\rho, \eta$ coordinates, the $A_{k-1}$ singularity is located at $\rho=0$ and $\eta=N/k$, where the
    slope changes.
    In general,
     we will find that if the slope changes by $k$ units at some value of $\eta$, then we have an
    $A_{k-1}$ at that point.

    After studying these examples, we can now formulate the general rule. We will check that the
    following rules lead to regular geometries.
    The solutions are  regular,   or have only    $A_{k-1}$ singularities,
     as long as we consider
    line charge densities whose slopes are integers.
     The slope is always decreasing. The slope can only
    change at integer values of $\eta$.
     We will show that these conditions follow from charge quantization. Note that the line
     charge density is simply the value of $y$ at that point, see \wardtrick .

    An alternative way to think about these line charge densities is  the following.
    We choose a sequence of
    integers $\lambda_i$, $i=1,2,\cdots$, with the condition that
    $\lambda_{i} \leq \lambda_{i+1} \leq N$. In addition
    the curve has to be convex,  $ 2 \lambda_i - \lambda_{i-1} - \lambda_{i+1} \geq 0$.
    See \profiles\ for   particular examples.
     It turns out that we can think of the $\lambda_i$ as the ranks
    of the nodes of a quiver that is ending a chain see \profiles .
     Each time the slope changes we need to add some flavors. In the   spacetime solution,
     each time the slope changes by $k$ units we have an $A_{k-1}$ singularity. If it changes by
     only one unit, we have a smooth geometry. The change of slope at point $i$ is given by
     \eqn\changesl{
     k_i \equiv 2 \lambda_i - \lambda_{i-1} - \lambda_{i+1} =
     ( \lambda_i - \lambda_{i-1} ) - ( \lambda_{i+1} - \lambda_i)
     }
     Then we have a quiver with gauge group $\prod_i SU(\lambda_i ) $. The $i$th gauge group has matter in the
     bifundamental connecting that node to the neighboring nodes. In addition it has $k_i$ fundamental
     hypermultiplets. $k_i$ also labels the type of $A_{k_i-1}$ that we have at that point.
     More precisely, the relationship between the quiver and the punctures is as follows \Gaiotto .
     The quiver we have described contains a number of coupling constants, the coupling constants of
     the various gauge groups. These are realized as additional elementary punctures. Thus, the
     composite puncture is what is left over after we go to a strong coupling region, which corresponds
     to moving away all the elementary punctures, leaving the final composite puncture. In particular,
     if we have the configuration with $\lambda(i)=i$, for $i\leq N$ and $\lambda(i)= N$ for $i\geq N$ and we move
     away the punctures we are left with a smooth region with no punctures. In all other cases we are left with
     some geometric structure at the origin.

     \ifig\arcs{ In (a) we have plotted the line charge density $\lambda(\eta)$ as a function of $\eta$. In
     (b) we have the corresponding $\rho , \eta$ plane. We have displayed  a couple of four cycles we can use
     to measure fluxes, namely $\overline{AB}$ and $\overline{BD} $. In addition we have also displayed
     the $S^3$ that is shrinking when we go from the $A$ to the $C$ segment. This $S^3$ consists of the arc
     $\overline{AC}$ together with the circles $S^1_{A}$ and $S^1_C $ which are the circles that
     shrink on the segments $A$ and $C$ respectively.
 } {\epsfxsize4in\epsfbox{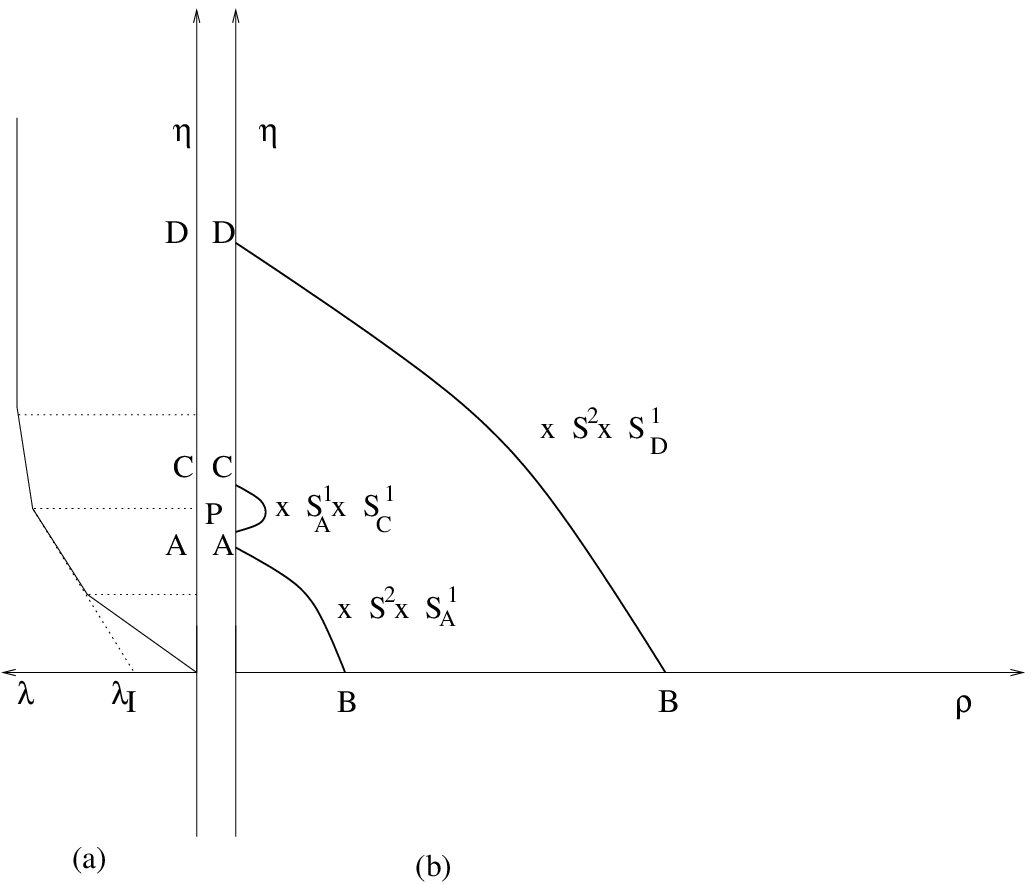} }

   We will be studying the solution \soltuc\ near $\rho \sim 0$ in a region where the slope of the charge
   density is constant.  In this region we can
   approximate the solution by
    \eqn\fullme{\eqalign{
  d s^2 \sim & \kappa^{2/3}  \left( {   \dot V \tilde \Delta \over 2 V'' } \right)^{1/3}
   \left[ 4 ds^2_{AdS_5} +   { 2  V'' \dot V \over \tilde \Delta} ds^2_{S^2} +
   {2  V''\over \dot V} ( d \rho^2 +  \rho^2  d\chi^2 + d\eta^2 )
   + \right.
   \cr
    & ~~~ + \left. { 4 \over   \tilde \Delta }
     ( d \beta +  \dot V'   d \chi )^2 \right]
\cr
\tilde \Delta \sim &    2 \dot V   V'' + ( \dot V')^2
\cr
C_3 \sim  & \kappa 2  \left[  ( - \dot V + \eta \dot V' ) d\chi +  ( { \dot V \dot V ' \over \tilde \Delta } - \eta ) ( d\beta
+ \dot V' d\chi )
\right] d\Omega_2
 }}
 where $\dot V' $ is the constant slope of the region we are considering.

   Now we will discuss some aspects of the topology and the charges of these solutions.
   This leads to a proof of the rules we enunciated above.
   At $\rho =0$ we see that the $\chi $ circle is shrinking.
    We also see that the circle given by $\beta + \dot V' \chi $ is not shrinking.
    Thus, we can define a new coordinate $\beta' = \beta + \dot V' \chi$ which is not shrinking.
    This makes sense only if the slope, $\dot V'$, is an integer.  This geometric condition is quantizing
    the slope and it is constraining it to be a constant along a segment in the  $\eta$ direction.
   The circle that is shrinking, together  with the two sphere, and an arc in the $\eta, \rho$ plane which goes
   from $\eta =0$ to the segment under consideration at $\rho=0$, see  the arc $\overline{AB}$ in \arcs,
   forms a closed four cycle.
   The flux of the four form field strength, $F_4$, on this four
    cycle is given by the intercept of the line going through the segment. This can be seen
    easily from \fullme\ by evaluating the difference in the $C$ field between the two ends of the arc
    $\overline{AB}$ in \arcs .  In other words, if the
    line charge density on the segment under consideration is $ \lambda =  r   \eta  + \lambda_I  $ then
    the charge is given by $Q_4 = \lambda_I$ (note that $\lambda_I \geq 0 $). Due to the way we constructed
    the segments $\lambda_I$ is always an integer.  We can interpret this charge as measuring the number
    of fivebranes that are creating the puncture. More precisely, in a situation where we have many segments,
    we consider the four cycle that is made out of the arc that goes from the $\eta =0$ plane to
    the last segment with a non-zero slope. This flux can be interpreted as the total number of M5 branes
    that we have combined to make the puncture.

     Let us consider, for example a simple configuration such as the one in \linecharges d, which
    can be interpreted as the backreacted solution of a number of fivebranes.
    Applying this construction to the segment $\overline {B D} $ in \arcs , gives us a charge $N$ which the
    total number of fivebranes  that we wrap on the hyperbolic space. On the other hand, if we consider an
    arc going from the $ \rho =0$ to the second segment in the \linecharges d , we get the number $K$ of
    M5 branes that are transverse to the original $N$ M5 branes.
    When we go back to the $y, x_1, x_2$ coordinates, this four cycle corresponds to the cups we displayed
    in \rods (b).

    Let us now study the solution near the point where two segments with different slope meet, such as
    the point $P$ in \arcs . Let us first notice that
  we  can   construct a three manifold from an arc going between two consecutive segments,
    see  $ \overline{AC}$ in \arcs , together with
     the two circles that are shrinking at each of the segments.
    This gives us a space which is topologically  $S^3/Z_{k}$ where $k$ is the change in slope that
    occurs when the two segments meet. It turns out that at this point
    we have an $A_{k-1}$ singularity in the geometry. If
    the change in slope is just one unit, we simply have an $S^3$ that shrinks smoothly at that point, as we
    had in the first two examples we discussed above. This fact can be seen by noting that
    $V''$ has a delta function source when the slope changes. This means that around this
    point we have
    \eqn\havecon{
     V''  = { k \over 2 }  { 1 \over \sqrt{ \rho^2 + ( \eta - \eta_0)^2 } }
     }
     when the slope changes by $k$ units.
     When we insert this in \fullme\ we see that the $\rho, \eta, \chi , \beta $ directions combine
     to give a space which is locally $R^4/Z_k$ at this point.
     Around this region we can approximate the metric as
        \eqn\annh{\eqalign{
  d s^2 = &  \kappa^{2/3} ( \dot V)^{2/3}
   \left[ 4 ds^2_{AdS_5} +    ds^2_{S^2} + ds^2_{R^4/Z_k} \right] + \cdots
   }}
   where $\dot V = \lambda(\eta_0)$ is the value of the line charge density where it changes slope.

    The $A_{k-1}$ singularity, for $k>1$, gives rise to non-abelian gauge fields in $AdS_5$ representing
    global symmetries of the theory. In the next subsection we discuss them in more detail.

     \ifig\twocycle{ We consider a linear quiver with increasing ranks (a) and the corresponding charge profile for
     the gravity solution (b).  The label $i$ denotes node number. $\lambda_i$ is the
      rank of the gauge group, $SU(\lambda_i)$,
     at each node. When there is a slope change we need extra fundamentals at the node. In (a), (b) the slope
     changes by one unit at $\eta = i$ and $\eta = i + 3$. We can form a BPS state by connecting these fundamentals
     using the bifundamentals   between these two nodes. In (c) and (d) we present a quiver and the corresponding
     charge profile. We consider a BPS state  $q_1 AAA q_3$,
      which contains a quark from the left node in (c) and bifundamentals
     that link it to the right node in (c). The corresponding cycle is a two sphere which is a bound state
     of the two two-spheres around each segment.
 } {\epsfxsize4in\epsfbox{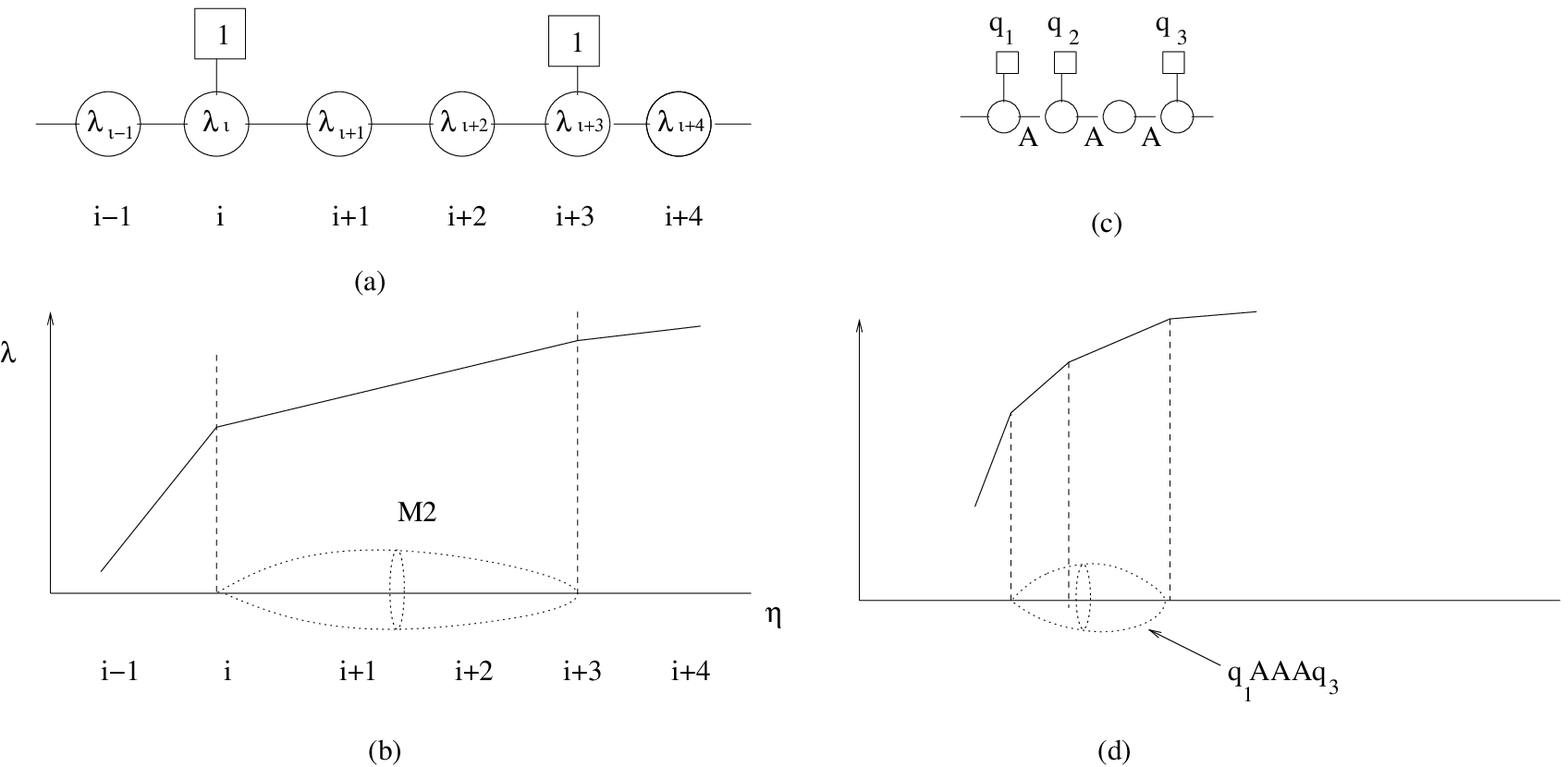} }

    In fact, even in the case that $k=1$ we can get a $U(1)$ gauge field in $AdS_5$.
    This can be seen as follows. First let us note that, given a segment, we can make a
    two cycle by considering the circle that does not shrink on the segment together with the segment itself.
    Since at the two ends the circle is shrinking, we see that we get a two cycle with the topology of $S^2$.
    The  $C_3$ gauge
    field with indices  on the two cycle gives rise to a gauge field in $AdS_5$ which is associated to
     a global $U(1)$ symmetry. The construction of
    these two cycles works for all the segments except for the first one, the one reaching $\eta =0$, since
    there the circle does not shrink.  Thus, if we have $n$ segments, we have $n-1$ $U(1)$ symmetries.
    In all cases, even for the first segment, we can form a four cycle by taking the
     segment, the circle that does not shrink and the $S^2$. The flux of the four form on this four cycle is
     given by $\eta_f - \eta_i$, the difference in $\eta$ between the two ends of the segments. Flux quantization
     implies that all the values of $\eta$ at which the slopes are changing are all integers.
     For all segments except the first,
     we can wrap an M2 brane on the segment and the circle that does not shrink.
     This M2 brane is a BPS state. It carries one unit of charge for the $U(1)$ symmetry
     associated to the
     $C_3$ field on this two cycle.  The flux on the corresponding fourcycle implies that
     the conformal weight of the state is given by
    $\Delta =  \eta_f - \eta_i$. Up to corrections of order one, this agrees with the conformal dimension of
    the following BPS state. Let us consider the point at the beginning of the interval. Let us say that
    this occurs at $\eta = i$. Let us say that the slope changes by one unit at this point. Then there is
    an extra fundamental at this point. Similarly, if the interval with constant slope ends at $\eta = j$, $i< j$,
    then there is at least one fundamental at this point. Then we can form an operator of the form
    $q_i A_{i,i+1} \cdots A_{j-1,j} q_{j}$, see figure \twocycle (a).  The total conformal weight of this state is
    $\Delta = j-i +2$. The $+2$ comes from the quantization of the fermion zero modes on the surface. This
    M2 brane is a two sphere and it has no other zero modes. In fact, in the limit that $j=i$, we get zero
    area and the field is naively massless. In fact, in this case we get an $A_{k-1}$ singularity, with $k\geq 2$,
    and the field is simply the partner of the extra non-abelian currents that we have at the $A_{k-1}$
    singularity. This field has $SU(2)_R$ spin one and thus has $\Delta =2$. This is an alternative
    explanation for the origin of the $+2$ in the formulas for the $SU(2)_R$ spin. In this case the
    origin of this spin comes from the fermion zero modes that are scalars on the worldsheet (after twisting).
   This implies that these zero mode scalars always have the effect of adding a $+2$ to the spin of the brane.
    Note that the $U(1)$ gauge field in $AdS_5$ coming from $C_3$ on the two cycle
     is simply given by the difference of the $U(1)$ flavor
    symmetry of the two extra fundamental hypers in \twocycle (a).

    Finally, if we have various segments with change in slope $k_i$ at various points, then the total global
    symmetry is simply $ ( \prod_i U(k_i) )/U(1) $. The overall $U(1)$ factor is not a real global symmetry of
    the system, and that is why we are modding out by it. Note that in this case we can construct a BPS
    state that starts with a quark at one place where the slope changes and we can go to any other place where
    the slope changes. In this case the total dimension of the state is $\Delta = \eta_f - \eta_i +2$. There
    is again a single $+2$ due to the fermion zero modes and nothing else. For example, in the case
    shown in \twocycle (c), the conformal weight of the state would be $3+2$ where 3 is the number
    of steps we move to the right. This state can be viewed as a bound state of two M2 branes wrapping the
    two two-spheres associated to the two segments in \twocycle (d).

  \ifig\multirod{  In (a) we see a particular charge profile that we discussed in the construction of
  $U(1)$ symmetric solutions, see \profiles . In (b) we see the corresponding charge distribution for the Toda equation
  that describes the puncture. We can use this to construct non-U(1) invariant solutions of the Toda equation
  involving this puncture.
 } {\epsfxsize4in\epsfbox{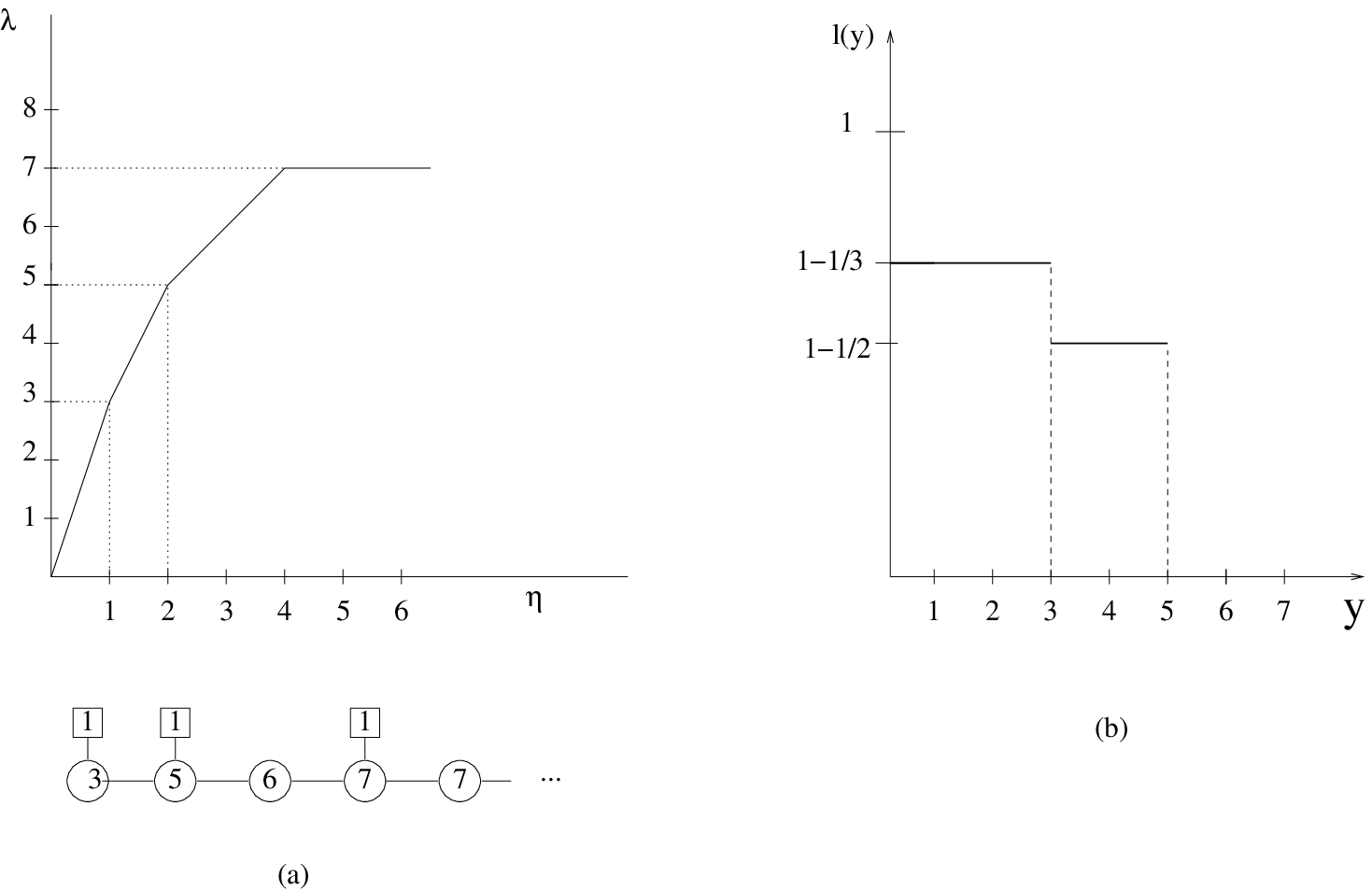} }

 Finally, we should mention that this detailed analysis of the solution with $U(1)$ symmetry can be translated
 into the corresponding boundary conditions for the Toda equation at the point that the puncture is inserted.
 Namely, if we consider the solution near a segment with constant slope $\dot V '$, then the potential behaves
 as $ V \sim \dot V(\eta ) \log \rho $. Then \wardtrick\ implies that
 \eqn\implic{
  \log r = \partial_\eta V = \dot V' \log \rho ~,~~~~~~~~
   \log D = 2 ( \log \rho - \log r ) = - 2 ( 1 - { 1 \over \dot V' } ) \log r
   }
   This expression for the charge density is correct for the range of $y$ where the slope is constant.
   Thus the final boundary condition for the Toda equation is of the form
   \eqn\todacond{
   \nabla D = - 4 \pi   \ell(y)
   }
   where $\ell(y)$ is a piecewise constant function which only changes at integer values of $y$. These constants
   decrease as we increase $y$. Finally, each of the constants takes a value $(1- 1/n)$ where $n$ is an integer
   (equal to $\dot V'$). See   \multirod (b). We can use these boundary conditions for the Toda equation
   to construct   solutions  with this  puncture which are not necessarily $U(1)$ invariant.

   Thus, we have given   all the information that is necessary to construct solutions associated to $N$ branes
   on a Riemann surface with any number of punctures.

      \ifig\sphereprofiles{  (a) We see the charge distribution which would be describing two punctures on
      the sphere with $K$ additional ordinary punctures where the ordinary punctures
      are smeared on the equator in a $U(1)$ invariant fashion.  In (b) we see the situation with $K = 2N$ and we see
      the emergence of an $A_1$ singularity, as we mentioned in the text. These profiles only tell us about
      the behavior of the solution near $\rho =0$. At large $\rho$ the solution should be replaced by a full
      non-U(1) invariant solution of Toda. In (c) we see the profile corresponding to a quiver
     with $K-1$ nodes. In (d) we see the extreme case where we have a single node.
 } {\epsfxsize4in\epsfbox{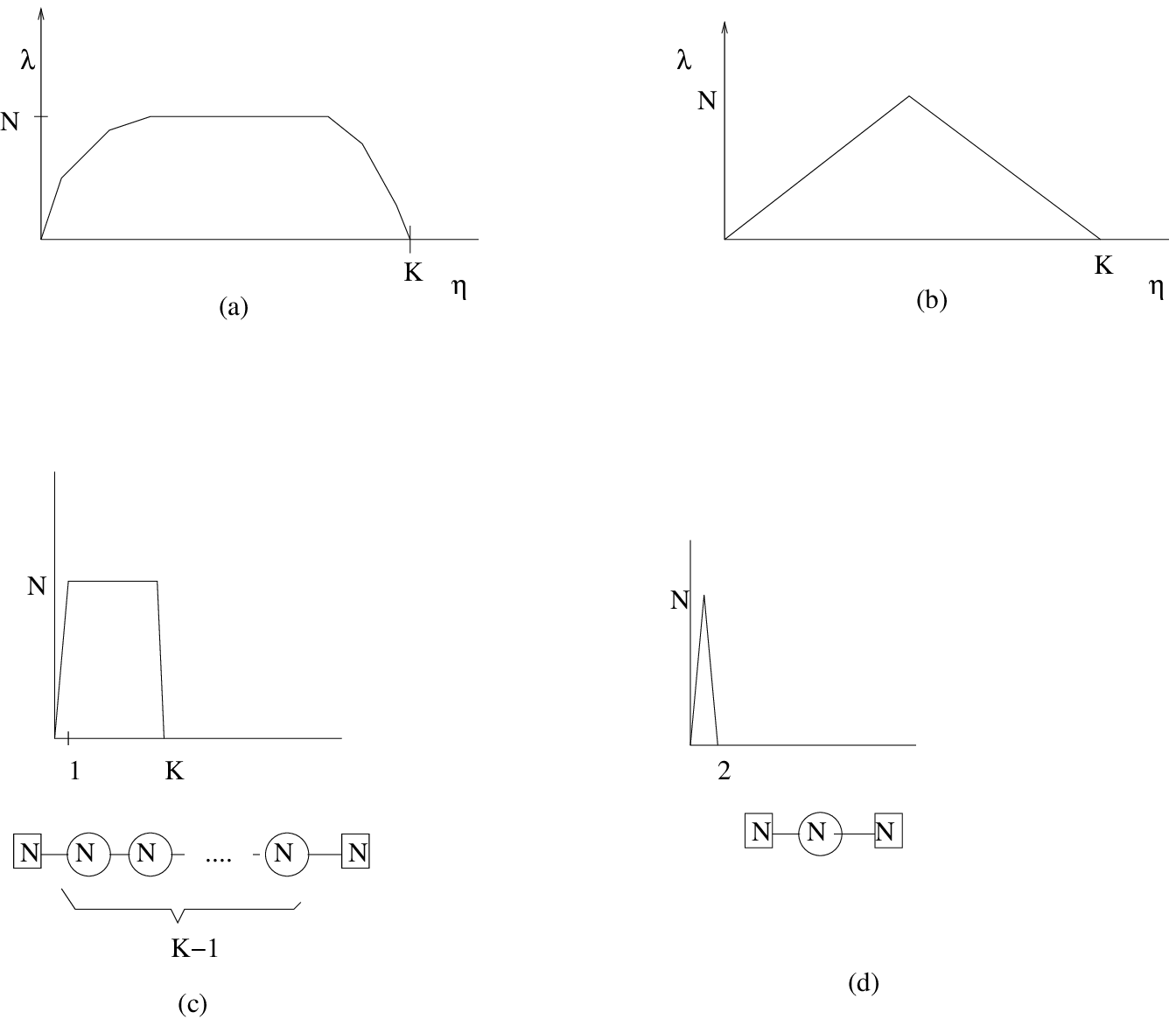} }

 Finally, we should comment that it is possible to consider some solutions which are $U(1)$ invariant that
 correspond to a sphere with two special punctures and many elementary punctures which are uniformly
 smeared along the equator of the sphere. This is a solution which will be singular on this equator
 but is regular everywhere else.
  These are given by charge distribution profiles which rise and then
 decrease again, such as the ones shown in \sphereprofiles . In the case that we have $K = 2 N$ ordinary
  punctures
 we can have a solution with an $A_1$ singularity in the interior, which is closely related to the
 solution we discussed near \approxso , see \sphereprofiles (b).
 Notice that the state we discussed near \totalv\ corresponds to a
 two cycle that is constructed from the horizontal segment in figure \sphereprofiles (a) and the circle that
 does not shrink there.

 Another interesting case one can consider a sphere with two $Z_N$ punctures, each with an  with $SU(N)$, symmetry
 and a bunch of ordinary punctures, see \sphereprofiles (c).
 As we decrease the number of ordinary punctures we eventually get to the situation displayed in
 \sphereprofiles (d). In this case we get an $A_{2N-1}$ singularity. This corresponds to the $SU(N)$ theory with
 $2 N $ flavors\foot{We have learnt that J. Polchinski and G. Torroba have
  been independently
 studying the gravity description of this theory.}.
  This does not give the full gravity dual, which  would require solving the full Toda equation
 with the two elementary punctures set at appropriate locations on the two sphere, rather than smearing them.
  This should be viewed as an  argument for
 the emergence of the $SU(2N)$ global
 symmetry from the bulk point of view.   It would be nice
 to analyze this case in more detail. One can check that the solution contains a small circle and one can reduce
 to type IIA and obtain a solution with string scale curvature.
  It is interesting that one can   view
   this solution as a  limit of the constructions in this paper.

 \newsec{ Some comments on solutions with   $A_{k-1}$ singularities }

 In this section we discuss solutions which contain $A_{k-1}$ singularities.
 Just to be concrete, let us consider a simple configuration with an $A_{k-1}$ singularity.
 We consider a profile with slope $k+1$ from $\eta =0$ to $\eta_k$ and then we continue with
 slope one until we get to $\lambda = N$. The final profile looks similar to the one in
 \linecharges (d).

 On an $A_{k-1}$ singularity in flat eleven dimensional space we have a seven dimensional
 $SU(k)$ gauge theory at low energies.
 In our case, the $A_{k-1}$ singularity is extended along $AdS_5$ and the two sphere, see \annh .
  It is interesting to compute the five dimensional gauge coupling of these fields.
    In seven dimensions the gauge coupling is  $ g^2_{7} = 32 \pi^4 l_p^3  $.
    This theory  wraps the $S^2$
    and is   reduced to five dimensions. We see from  \annh\  that
    the area of the $S^2$ is given by the value of $\lambda$, or $y$,
     where the
    change in slope is happening. In other words, it is given by the $\lambda(\eta_k)$ where the $A_{k-1}$ is living.
    Taking into account all factors we find that the five dimensional gauge coupling is
    \eqn\fivedg{
    { R_{AdS_5} \over g_5^2 } = { 1 \over 8 \pi^2 }~ { \lambda(\eta_k) }
    }
    This value of the gauge coupling determines the two point function of the $SU(k)$
     global symmetry currents. It is
    also related the mixed anomaly involving two global currents and the $U(1)_R$ current.
    We can also view this number as follows. Imagine that we were to weakly gauge the $SU(k)$ gauge symmetry.
    Then the ${\cal N}=2$ theory would contribute to the running of the coupling as
    \eqn\runncu{
    { 8 \pi^2  \over g_4^2 } = - 8 \pi^2 { R_{AdS} \over g_5}   \log( \rm{Scale} )=
     - { \lambda(\eta_k) } \log( \rm{Scale} )
    }
     We can compare this to the running we would get from
   $n_h$ hypermultiplets   ${ 8 \pi^2 \over g_4^2 } = -   n_h  \log(\rm{Scale} ) $.
    Comparing this to \fivedg\ we see that the
   running is the same as that of $n_h = \lambda(\eta_k)$ hypermultiplets. But $\lambda(\eta_k)$ is
   precisely the rank of the gauge group associated to the node $\eta_k$. Thus, the final result is
   precisely as we would expect from the correspondence between punctures and quivers as summarized
   in  \profiles . Of course, we get agreement due to the fact that  this
   coefficient is related by supersymmetry
   to an  anomaly and is   thus   independent of the coupling.
   Therefore, when we start with the quivers in
   \profiles\ and take the couplings to infinity, the anomaly remain the same. Of course, this is a check
   that we have made the correct identification between the different punctures and the corresponding geometric
   structures, which are determined by the charge density profile.

 Another interesting quantity is the contribution of the $A_{k-1}$ singularity to $a-c$.
 Such terms must arise from higher curvature contributions to the action. In particular, on the
 $A_{k-1}$ there is an $R^2$ correction which   descends to an effective $R^2$ correction to the
 action in $AdS_5$. As we discussed above, this leads to a contribution to $a-c$ \refs{\AharonyRZ,\BlauVZ,\NojiriMH}.
 In order to compute the coefficient of this contribution we note that we can start with $k$ D6 branes in
 type IIA theory, where this coefficient has been computed in \BachasUM . After lifting those formulas
 to M theory we find that the relevant contribution is\foot{ The overall sign is defined so that
    that in Euclidean
 space the partition function is $e^S $.}\foot{
 This really the contribution of $k$ Kaluza Klein monopoles. On $A_{k-1}$ singularities we need to replace
 $k \to k - { 1 \over k } $. In principle, we would need to add the contribution of the $A_{k-1}$ singularity and
 the contribution from the smooth geometry around it. We have not computed the latter in detail, but we are
 going to see that treating them as KK monopoles one seems to capture correctly that geometric piece.  }
 \eqn\resan{
S = k { 1 \over ( 2 \pi)^4} { 1 \over  3 . 2^6 l_p^3 } \int d^7 x   R_{\mu \nu \delta \rho} R^{ \mu \nu \delta \rho}   + \cdots
}
where the dots are terms which are irrelevant for our computation.
We are wrapping
this over a sphere of size give by $\lambda(\eta_1)$. After reducing on this sphere we get an effective five
dimensional term of the form
\eqn\fivedcontr{
S = k \, \lambda(\eta) \,   { 1 \over ( 2 \pi)^2} { 1 \over  3 . 2^6   } \int_{AdS_5}  d^5 x
 R_{\mu \nu \delta \rho} R^{ \mu \nu \delta \rho} + \cdots ~,~~~~~~~~~~R_{AdS_5} =1
}
where the radius of $AdS_5$ is one.
On the other hand, for small $|c-a| \ll a,c$ we have that the corresponding correction should be
\refs{\NojiriMH,\Kats}
\eqn\computs{
S = { (c-a) \over 16 \pi^2 } \int_{AdS_5}    R_{\mu \nu \delta \rho} R^{ \mu \nu \delta \rho} + \cdots
  ~,~~~~~~~~~~R_{AdS_5} =1
}
Comparing \computs\ and \fivedcontr\ we can read off
\eqn\contribf{
n_v - n_h = 24 ( a-c) =  - { 1 \over 2 } k \lambda(\eta_1)
}

 This should be compared with the field theory computation of $a-c$.
 When we consider a field theory
 quiver given by a profile $\lambda(\eta)$, the total contribution to $a-c$ is given by\foot{ This sum is sensitive
 to how we cut if off. We are summing in such a way that the index $i$ runs over each node and when
 we have a bifundamental, we assign half of its contribution to each of the nodes that it connects. Then
 we cut off  by summing up to a maximum node.}
 \eqn\contrib{
 24 ( a- c) |_{quiver}  = n_v - n_h =    - \sum_{ i\geq 1 } 1 - { 1 \over 2 } \sum_{ i=1}k_i \lambda_i
 }
 So we see that at each point $\eta = i$ where the slope changes by $k_i$ units there is a contribution
 proportional to $k_i \lambda_i/2$. The first term,  the sum  of ones,  is equal to the
  number of
  elementary punctures. More precisely,
 the quiver given by the nodes $i$, $\lambda_i$, corresponds to the composite puncture whose gravity dual
 we are analyzing
 plus a number of elementary punctures which is equal to the number of nodes of the quiver. When we isolated
 the composite puncture we moved away the elementary punctures. Since the first term is proportional to the
 number of punctures, we can associate it to the punctures we moved away. As a consistency check, notice
 that if we have the profile corresponding to a single elementary puncture, which has
 $\lambda_i = 1 + i $, for $i\geq 1$ up to $\lambda = N$. This has $k_1=1$ and $\lambda_1=2$. Thus the second
 term in \contrib\ indeed gives a minus one.

 \ifig\simplecases{ Two simple cases that give rise to $SU(k)$ global symmetry.
 (a) We start with a segment of slope $k+1$ ending at $\eta=1$ and then we continue with a slope 1 until we reach
 $\lambda = N$. In (b) we have a segment of slope $k$ and it ends at $\eta = N/k$. This is possible if $N$ is a
 multiple of $k$.
 } {\epsfxsize4in\epsfbox{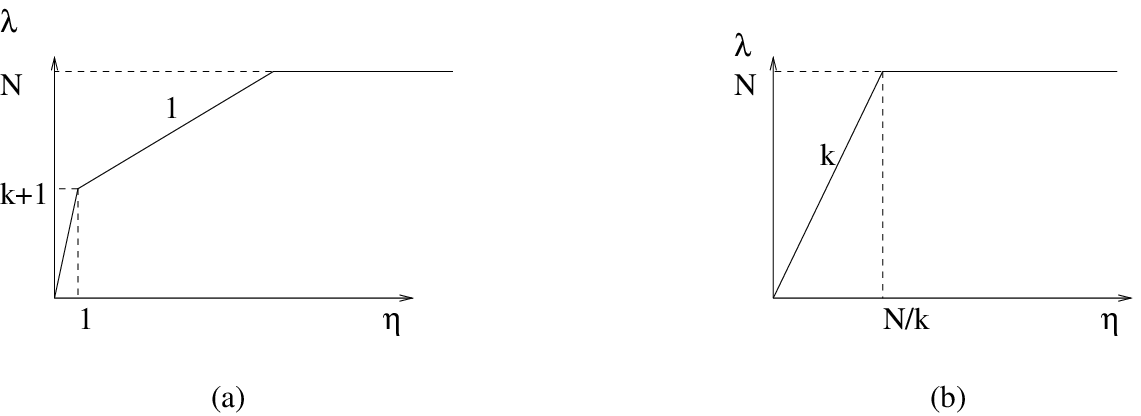} }

   One simple case with an  $SU(k)$ symmetry is  $\lambda_1 = k+1 $, and then $\lambda_i = k+i$ for $i\leq N-k$,
   see \simplecases (a).
   This corresponds to ungauging one of the gauge group factors in the
   chain of increasing gauge groups in \quiver (b) . This gives a configuration with $SU(k)$ symmetry with the
   lowest value for the two point function of the $SU(k)$ global currents, namely,
    a contribution equal to that of $k+1$ hypers of $SU(k)$. This is the lowest value within the class of theories
    we are considering in this paper. It would be nice to understand if there is a strict lower bound.

   A second possibility arises in the case that $N$ is a multiple of $k$. In such a situation
  we can have chain with slope $k$ which goes all the way to $N$, see figure \simplecases (b).
   Then at $\eta = N/k$ the slope
  changes from $k$ to zero and we have an $A_{k-1}$ singularity. This is simply a $Z_k$ quotient of
  the hyperbolic space solution. This gives the highest contribution to the two point function for the $SU(k)$
  currents, namely, equal to that of $N$ hypers of $SU(k)$. Note that in this case the singularity
  is located at $\eta = N/k$ and not at $\eta =0$. In fact, at $y=0$ the solution is completely
  smooth. More explicitly, if we perform the $Z_k$ quotient of the hyperbolic space solution \metrieleven ,
  we are quotienting under $\beta \to \beta + 2 \pi/k$ and $\chi \to \chi + 2 \pi/k$. Thus, at $r=0$, the
  solution is only singular when the $\chi$ circle is also shrinking, which happens at $y=N$.
  It is interesting that in this case, the circle at $\eta =0$ that is not shrinking has a size
  which is reduced by an amount of order $k$. More precisely the geometry near $\eta \sim 0$, $\rho \sim 0$
  is given by
  \eqn\geoemsim{
  ds^2 = ( {\rm factor} ) \left[ 4ds^2_{AdS_5}+ { 4 \over k^2 } d   {\beta'}^2 + d \tilde \eta^2 + \tilde \eta^2 d\Omega_2^2
  + d\tilde \rho^2 + \tilde \rho^2 d\chi^2   \right]
}
where $ \beta' = \beta + k \chi$ and the tilde variables are a simple rescaling of the un-tilded ones.
Thus, for large $k$ we can envision a reduction to a type IIA theory, at least near this region.

  Note that as $k \to N$ the two above simple cases, \simplecases ,
   merge and there is only one puncture
   with $SU(N)$ global symmetry:  the $Z_N$ quotient
  of hyperbolic space. This is the special puncture which we
  can use to construct the theory $T_N$. The theory $T_N$ consists of a sphere with three such punctures. In the
  next section we describe more explicitly the gravity solution corresponding to this case.

   \ifig\mfivebranes{ (a) We start with a geometry containing a segment of slope $k$ that starts from the
   origin. Eventually this segment will end when it reaches $\lambda =N$, but we focus on the
   region near $\eta \sim 0$. Near $\eta \sim 0$ we have a circle that does not shrink. We wrap
   $M$ M-fivebranes on this circle. There are various charge configurations which could describe
   this system of fivebranes, such as (b) or (c). These corresponds to branes that have blown up in various ways
    due
   to the Myers effect.
 } {\epsfxsize4in\epsfbox{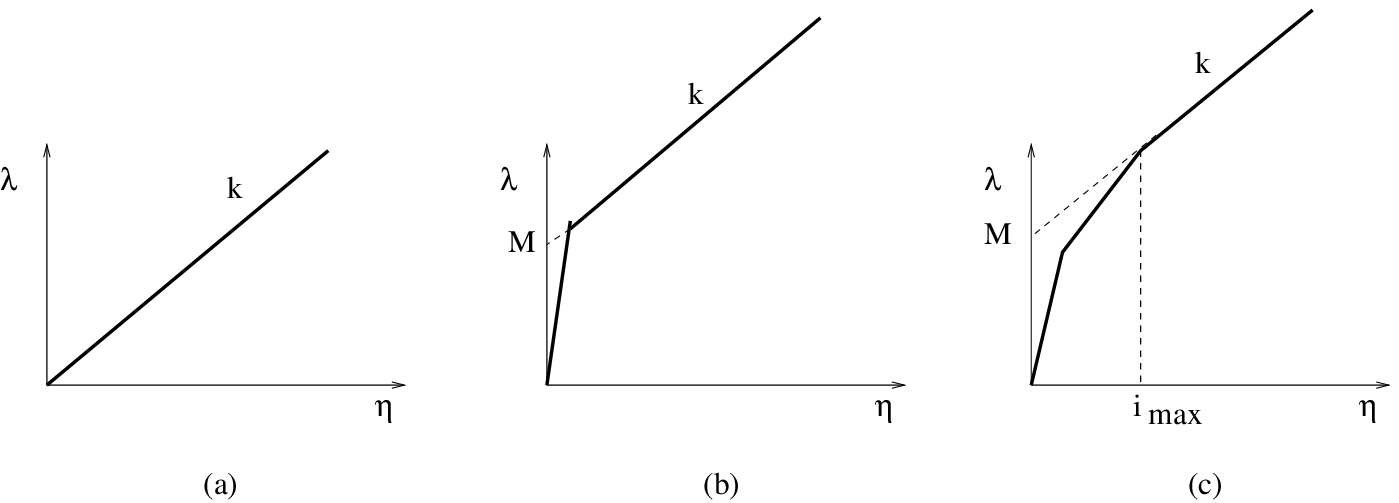} }

  Finally, let us discuss some qualitative aspects of these solutions. Let us discuss the different
  solutions we can make with $M$ fivebranes. More concretely, let us start with a background which
  contains only a segment of slope $k$ as we discussed above, see   \mfivebranes (a).
   This background contains an $AdS_5 \times S^1$ near
  $\eta = \rho =0$, see \geoemsim .
   We now add $M$ fivebranes wrapping this circle. We already know the full backreacted
  description of these geometries. We have to consider charge profiles which end up with the segment of slope
  $k$, as shown in   \mfivebranes (b),(c). We would now like to answer the question: Why is it that $M$ fivebranes
  can form these various states?.
   This can be understood more readily in the large $k$ limit. In this case the circle in \geoemsim\ is getting
   small and we can reduce type IIA. The 5-branes become $M$ D4 branes. The crucial fact is that
   these D4 branes are subject to a transverse $H_3^{NS}$ field. Then the Myers effect implies that
   the D4 branes can blow up into D6 branes \myers\ on a fuzzy $S^2$. Namely, we can view the
   D6 branes as arising from the various ways we can construct an $SU(2)$ representation of dimension $M$.
   These are given by the partitions of $M$.
   Indeed the various possible profiles are also labeled by the partitions of $M$. We can make this
   manifest as follows. The profile is given by a sequence $\lambda_i$ which eventually becomes the line
   $\lambda_i = M + k i $ for $i\geq i_{max}$, see \mfivebranes (c).
   Let us define $\hat \lambda_i =  \lambda_i - k i  $ and
   $s_i = \hat \lambda_i - \hat \lambda_{i-1}$. Note that $s_i =0$ for $i> i_{max}$. Note also that
   $\sum_i s_i = M$. We now further define $n_i = s_i - s_{i+1}$.
   We the see that $M = \sum_{i=1} i n_i $. Thus we interpret $n_i$ as the number of SU(2) representations
   of dimension $i$.
   For example, in the case in \mfivebranes (b) we have that $n_1 = M$. This is the case where the $D4$ branes
   were not blown up and we have an unbroken $SU(M)$   symmetry. Another extreme case is   $n_i=0$ except for
     $n_M=1$. In this case the D4 branes have blown up into a single D6 brane.
     When we go back to M-theory the D6 branes become KK monopoles, and coincident
      D6 branes are $A_n$ singularities.

    This description is good if $k$ is large, but by a naive extrapolation, we also get the right qualitative
    picture even for $k =1$.

\newsec{ Simple quotients of hyperbolic space }

 There is a simple geometric construction involving three $A_{k-1}$ singularities.
   It is a Riemann surface with constant negative curvature with three such cusps.

   \ifig\pants{In (a) we display the fundamental region of hyperbolic space with three $Z_k$ singularities.
   These are placed at the three points $A,~B,~C$. A $2 \pi/k$ rotation around $A$ maps $B$ to its image.
   The fundamental region consists of both hyperbolic triangles depicted here.  The marked lines are identified.
   In (b) we show the large $k$ limit
   of $(a)$. We have a surface with three cusps placed at $A, ~B,~C$. Now these points are at the boundary
   of $H_2$.  In $(c)$ we have a more artistic representation of the surface in (b) as a pants diagram with three
   asymptotic regions.
 } {\epsfxsize4in\epsfbox{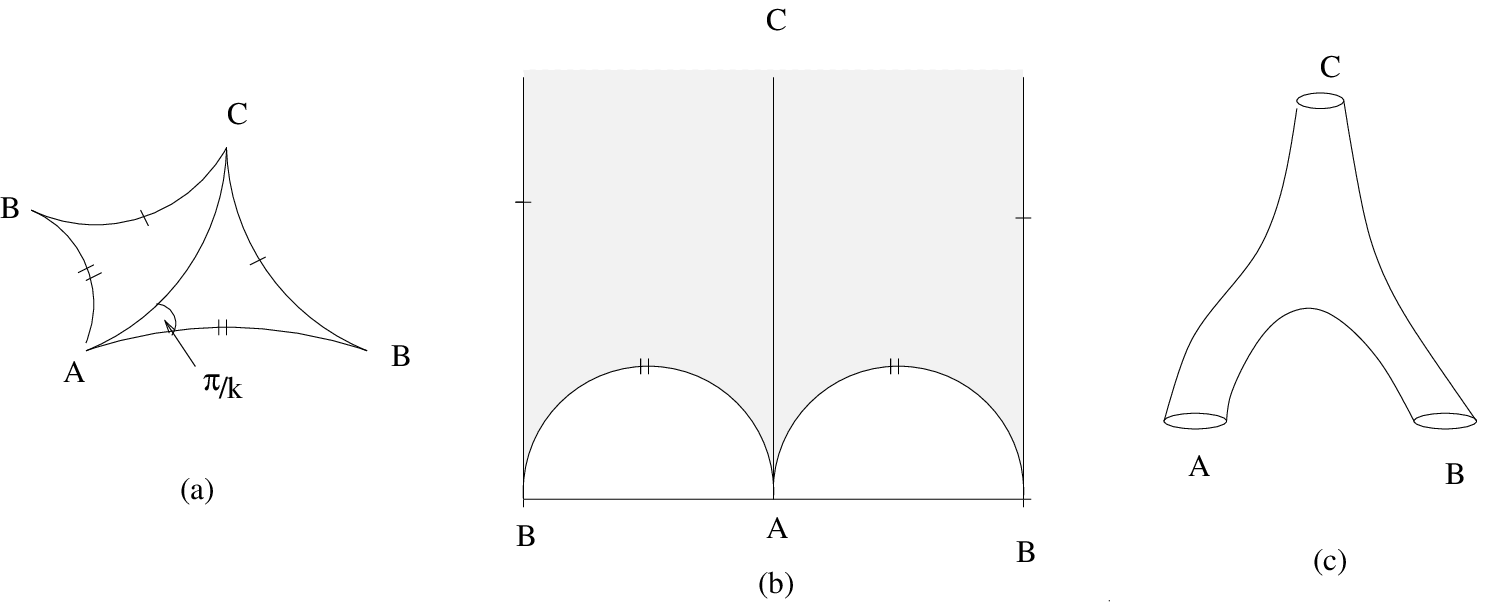} }

Let us describe this case very explicitly. We can view the upper half plane as
$SL(2)/U(1)$. We consider equivalence classes given by $g \sim g e^{ i \alpha \sigma_2 } $, $g \in SL(2)$.
We
denote the corresponding class by $[g]$. We can then represent a point on $H_2$ as
$[ e^{ i \beta { \sigma_2 \over 2} } e^{ \rho { \sigma_3 \over 2 } } ]$.
With this choice the metric in hyperbolic space is $ds^2 = d\rho^2 + \rho^2 d \beta^2 $.
We then see that if we set one of the singularities at the origin, say $A =  [ 1 ] $. Then
the other two points are at $B= [ e^{\rho { \sigma_3 \over 2} } ]$ and $C=[ e^{ i { \pi \over  k} { \sigma_2 \over 2 }} e^{\rho { \sigma_3 \over 2} } ]$, see \pants (c).
The requirement that these choices are consistent with a $Z_k$ symmetry at each point implies that
\eqn\implic{
\cosh { \rho \over 2} = { 1 \over 2 \sin { \pi \over 2 k } }
}
Notice that $\rho$ is the distance between any two orbifold singularities. Note that the structure
of this Riemann surface is completely rigid, it does not have moduli.
  We can   compute the area of this Riemann surface
  \eqn\areacom{
  { A_{\Sigma} \over 4 \pi } = { { 1 \over 2 }  - { 3 \over 2 k } }
  }
  Note that for $k=3$ we have that $\rho =0$ and we have a zero area surface. This is
   simply saying that we cannot embed a usual equilateral triangle in negatively curved space.
   For $k>3$ the surface
  has a non-zero area and $\rho > 0$.
  The purely geometric contributions to the central charges can be parametrized as
  \eqn\purec{ \eqalign{
  n_v|_{geom}  = &  (  { 1 \over 2 }  - { 3 \over 2 k } ) ( { 4 N^3 \over 3 } - { N\over 3 } )
  \cr
  n_v - n_h |_{geom}  = & (  { 1 \over 2 }  - { 3 \over 2 k } ) N
  }}
  In addition we should add the contribution from the $A_{k-1}$ singularities.
  These will contribute to $n_v$ and also to $n_v-n_h$. The contribution to $n_v-n_h$ is
  given by $ - (k -1/k) N/2$ for each of the three $A_{k-1}$ singularities.
  Finally, we conclude that the total contribution is
  \eqn\total{
  n_v - n_h = - { 3 \over 2 } ( k - { 1 \over k } )  N + (  { 1 \over 2 }  - { 3 \over 2 k } ) N =
  - { 3 k N \over 2 } + { N\over 2}
  }
 We have not computed the contribution of the $A_{k-1}$ singularity to $n_v$.

  An interesting special case is $k=N$. In this case we have the geometry dual to the
  theory $T_N$. We see that $n_v - n_h = -{ 3 N^2 \over 2 } + { N\over 2} $. This agrees
  with the field theory computation \tnval\ up to the $+1$ which could come from the
  center of mass hypermultiplet. It would be nice to match also
  this term of order one. Notice that the first two terms come from subleading corrections in $AdS_5$. The
  first, which goes as $N^2$, comes from the contribution at the $A_{N-1}$ singularities and the
  second comes from the $R^4$ correction in the bulk of the surface.
  In this case we see clearly the three $SU(N)$ gauge symmetries that arise on the three $A_{N-1}$
  singularities of the geometry.
  Notice that close to the singularities we have a circle that is shrinking to zero. In fact, since
  we have $N$ M5 branes wrapping this shrinking circle, we can go to type IIA string theory and obtain
  $N$ D4 branes. The gauge coupling on the D4 brane is becoming very weak, which explains in another way
  the origin of the $SU(N)$ gauge symmetry.

  Notice that we can formally send  $k \to \infty$. In this case the three $Z_k$ singularities go to the
  boundary of hyperbolic space and we end up with a solution with three cusps, as shown in
  \pants (a). The metric near each  cusp looks like the large $y$ region of
  ${ dx^2 + dy^2 \over y^2}$ with $x\sim x+2$, see \pants (b).
 This Riemann surface can be viewed as a pants diagram, see \pants (c). This is an approximate description
 for the surface at large $k$. In our solutions the maximum value of $k$ is $N$. But $N$ can be large so this
 approximate description becomes good for large $N$.

\subsec{ Comment on possible phenomenological applications }

In this section we   want to make a simple comment regarding a possible application of these
results.

In some phenomenological constructions one   sometimes postulates  the existence of an extra hidden sector describing
a new set of particles. In some cases this extra sector has a global $SU(k)$ symmetry. This $SU(k)$ symmetry
can then be gauged when we couple the extra sector to the rest of the theory. For example, if one is
interested in a Technicolor-like explanation of $SU(2)$ symmetry breaking, then one can consider a
theory with a global $SU(2)$ symmetry which is then gauged. Other applications for hidden sectors include
models for susy breaking, dark matter, inflation, etc.
In such constructions one would typically like to ensure that the contribution of the extra sector to the
running of the $SU(k)$ gauge coupling is not too large, in order to avoid the so called Landau poles.

A calculable class of theories are those with gravity duals. If the dual is a string theory, such as the
ones that would arise from taking a 't Hooft-like limit of a large $N$ field theory, then one typically
finds that the current two point function for the $SU(k)$ global symmetry scales with $N$.
For example, we can consider a QCD-like theory with $N$ colors and $k$ flavors. In the bulk, this scaling
with $N$ is due to the fact that flavor symmetries are realized on  D-branes . The D-brane
action has a bulk gauge coupling for $SU(k)$ which is $1/g^2 \sim 1/g_s \sim N$.
It is important that this factor of $N$ comes from the branes and does not come from the fact that we
have a large radius geometry.

One would like to alleviate this problem by considering situations where the five dimensional gauge coupling
in the bulk is large. But, how large can we make it?.

The theories we have described in this paper can give  rise to $SU(k)$ global symmetries with two point functions
which do not grow with $N$. Thus, the full theory can have a weakly curved gravity dual but nevertheless have a
global symmetry with an order one two point function. This arises because the $SU(k)$ global symmetry lives on
M-theory objects, such as M5 branes or KK monopoles. In other words, we get an $SU(k)$ symmetry from $k$ M5 branes
wrapping $AdS_5 \times S_1$. If the radius of $S^1$ were small, we would get a five dimensional gauge theory
in $AdS_5$ with a coupling proportional to $R_{AdS_5}/g_5^2 \sim { R_{AdS_5} \over R_{S^1} } $. Thus, when the radius
of the $S^1$ is comparable to the radius of $AdS_5$ we can have current two point functions
 of order one. Of course, in
that case, the theory on the brane is becoming strongly coupled. Nevertheless, in these ${\cal N}=2$ cases we
can compute the two point function either from the field theory quiver, or from the mixed anomaly.
More precisely, we see from \runncu\ that the contribution to the beta function is that of $\lambda(n_k)$
hypermultiplets. In the class of geometries we considered
this can be as low as $k+1$ and does not grow with $N$.

Of course, for phenomenological applications one would like to consider theories with less supersymmetry,
theories which are not conformal, etc. It seems possible to consider variations of this construction which
would contain well understood geometries like the Klebanov Strassler one \KS . We leave a full analysis of those
cases to the future.

\newsec{ Dicussion and conclusions }

In this paper we have  explained how to find the gravity duals of the large class of
 ${\cal N} =2$ superconformal field
theories discussed in
 \Gaiotto .
  They are found by solving a certain Toda equation with appropriate boundary conditions.
 The problem mirrors the description in the field theory. One needs to consider the Toda equation on a
 Riemann surface with some punctures. There is a rich set of possible punctures.
 We described some explicit properties of the solutions near the punctures.
 We also discussed solutions with no punctures. Such solutions simply involve a negatively curved Riemann surface.
 We have matched the leading and subleading contributions to the central charges $a,c$.
 In this regard, notice that the theories discussed here can produce both positive or negative values of $a-c$.
 The gravity solutions we described here include cases where one naively was not guaranteed to find a gravity solution,
 such as the $SU(N)$ theory with $2N$ flavors.

  We have mentioned that these theories arise from wrapping M5 branes on Riemann surfaces. It would be nice
  to study the flows between the 6d theory and 4d fixed point in more detail, in order to understand how
  the non-uniformities of the metric of the 2d surface flow to the IR. We have also discussed the inverse
  process of constructing the six dimensional theory from a limit of four dimensional theories. This gives
  some idea about the origin of the $N^3$ number of degrees of freedom. The four dimensional theories
  have of the order of $N A_2$ SU(N) gauge factors, where $A_2$ is the area of the Riemann surface on which
  the (0,2) theory is wrapped, in units of the ``lattice spacing''.

 It would   be interesting to study other aspects of these solutions. For example, one could
   compute the subleading corrections to $c$ and $a$ individually for the various punctures from the gravity
 solutions. Here we have only matched $a-c$. One can also compute the metric on the space of couplings from the
 gravity side. It would be nice to understand whether this can also be computed from the field theory side.
 For example, for the theories based on smooth higher genus Riemann surfaces, we get the Weyl-Peterson metric
 in Teichmuller space.

 One class of interesting ${\cal N}=2$ superconformal field theories that we did not discuss are
 the Argyres Douglas fixed points. It would be nice to understand how to extend the gravity description to theories
 in that class. They appear to require a new kind of puncture. This intuition is based on the results of
 \GMNtwo . One could also extend this to  the D type (0,2) theories.

 It would be nice to extend this description to theories with ${\cal N}=1$ supersymmetry. In fact, in \jmcn\ solutions
 based on Riemann surfaces were also constructed with ${\cal N}=1$ supersymmetry.

 One would also like to extend this to three dimensional theories based on wrapping Riemann M5 branes on
 quotients of $H_3$, see \refs{\PerniciNW,\AcharyaMU,\GauntlettNG}.


 { \bf Acknowledgments }

We are very grateful to N. Arkani-Hamed, H. Lin, J. Polchinski,  Y. Tachikawa and G. Torroba  for discussions.
We also thank A. Zhiboedov for correcting a figure. 
 
This work   was  supported in part by U.S.~Department of Energy
grant \#DE-FG02-90ER40542.

 { \bf Note added }

\lref\FSads{
  A.~Fayyazuddin and D.~J.~Smith,
  JHEP {\bf 0010}, 023 (2000)
  [arXiv:hep-th/0006060].
}

\lref\FSfirst{
  A.~Fayyazuddin and D.~J.~Smith,
  JHEP {\bf 9904}, 030 (1999)
  [arXiv:hep-th/9902210].
}

Solutions describing the M5 brane on Riemann surfaces were proposed in \FSads\ (based on \FSfirst ).
However, the solutions in \FSads\ are simply a coordinate transformation of the $AdS_7 \times S^4$
solution. This can be seen as follows. The metric is given in eqn. (31) of \FSads .
Up to an overall constant in the metric  we can set $ \pi N =2$ in  eqn. (3) of  \FSads .
One then defines a new complex variable
 $\tilde u = { 1 \over 4 } \int^z { dz' \over F(z')} $.
 We define a new coordinate $ \theta_{here}$ via
 $ { \cos \theta_{here} \over \sin \theta_{here} } = { \cos^2 \theta_{FS} \over \sin^2 \theta_{FS} |F| } $,
 where $\theta_{FS}$ is the $\theta$ variable used in \FSads .
 Finally, we  define $r$ and $\beta$ through  $ r e^{ i \beta } = ( \tilde u)^2$.
 With these definitions the metric takes the form of the $AdS_7 \times S^4$ metric as written in \midad .
 The simplest way to check this is to start from \FSads , and write their metric in terms of the
 ansatz \ansatz . In that way one sees that $y^2 = { 4 \cos^2 \theta_{FS} \over \alpha^2 } $. One then
 subtracts from the metric in \FSads\ the $dy^2$ term that we have in the ansatz. After doing this
 the metric simplifies a great deal. After  introducing the $\tilde u$ variable as above,   one
  gets the $AdS_7 \times S^4$ metric as written in \midad \findres .
\listrefs

\bye